\documentclass[10pt]{iopart}
\usepackage{graphicx}
\usepackage{subcaption}

\expandafter\let\csname equation*\endcsname\relax
\expandafter\let\csname endequation*\endcsname\relax

\usepackage{amsmath}
\usepackage{mathtools}
\usepackage{array} % for table
\usepackage{xr} % for referencing suplemental figures
\usepackage{float} % for the [H] specifier
\usepackage{array} % for table

\usepackage{cite}

\begin{document}

\title[Enhancement of Photoresponse for InGaAs Infrared Photodetectors]{Enhancement of Photoresponse for InGaAs Infrared Photodetectors Using Plasmonic WO$_{3-x}$/Cs$_y$WO$_{3-x}$ Nanocrystals}

\author{Z. D. Merino$^{1, 2}$, Gy. Jaics$^{3}$, A. W. M. Jordan$^{1}$, A. Shetty$^{1}$, P. Yin$^{3}$,  M. C. Tam$^{1, 4, 5}$, X. Wang$^{1, 3}$, Z. R. Wasilewski$^{1, 2, 4, 5, 6}$, P. V. Radovanovic$^{3,5}$, J. Baugh$^{1, 2, 3, 5, 6}$\footnote{Author to whom any correspondence should be addressed.}}

\address{$^{\text{1)}}$ Institute for Quantum Computing, University of Waterloo, Waterloo N2L 3G1,
Canada}
\address{$^{\text{2)}}$ Department of Physics, University of Waterloo, Waterloo N2L 3G1, Canada}
\address{$^{\text{3)}}$ Department of Chemistry, University of Waterloo, Waterloo N2L 3G1, Canada}
\address{$^{\text{4)}}$ Department of Electrical and Computer Engineering, University of Waterloo, Waterloo N2L 3G1,
Canada}
\address{$^{\text{5)}}$ Waterloo Institute for Nanotechnology, University of Waterloo, Waterloo N2L 3G1,
Canada}
\address{$^{\text{6)}}$ Northern Quantum Lights Inc., Waterloo N2B 1N5, Canada}
\ead{zmerino@uwaterloo.ca, baugh@uwaterloo.ca}

\vspace{12pt}
% \begin{indented}
% \item[]March 2024
% \end{indented}
\begin{abstract}
Fast and accurate detection of light in the near-infrared (NIR) spectral range plays a crucial role in modern society, from alleviating speed and capacity bottlenecks in optical communications to enhancing the control and safety of autonomous vehicles through NIR imaging systems. Several technological platforms are currently under investigation to improve NIR photodetection, aiming to surpass the performance of established III-V semiconductor p-i-n (PIN) junction technology. These platforms include in situ-grown inorganic nanocrystals and nanowire arrays, as well as hybrid organic-inorganic materials such as graphene-perovskite heterostructures. However, challenges remain in nanocrystal and nanowire growth, large-area fabrication of high-quality 2D materials, and the fabrication of devices for practical applications. Here, we explore the potential for tailored semiconductor nanocrystals to enhance the responsivity of planar metal-semiconductor-metal (MSM) photodetectors. MSM technology offers ease of fabrication and fast response times compared to PIN detectors. We observe enhancement of the optical-to-electric conversion efficiency by up to a factor of $\sim$2.5 through the application of plasmonically-active semiconductor nanorods and nanocrystals. We present a protocol for synthesizing and rapidly testing the performance of non-stoichiometric tungsten oxide (WO$_{3-x}$) nanorods and cesium-doped tungsten oxide (Cs$_{y}$WO$_{3-x}$) hexagonal nanoprisms prepared in colloidal suspensions and drop-cast onto photodetector surfaces. The results demonstrate the potential for a cost-effective and scalable method exploiting tailored nanocrystals to improve the performance of NIR optoelectronic devices.
\end{abstract}

\vspace{2pc}
\noindent{\it Keywords}: NIR photodetectors, NIR plasmonics, MSM photodetectors, Nanocrystals, 
    Nanorods, WO$_{3-x}$, Cs$_{y}$WO$_{3-x}$, InGaAs, InAlAs
%

\maketitle

\section{Introduction}

Light detection in the near-infrared (NIR) and infrared (IR) wavelength ranges has numerous applications in fields such as biomedical imaging and sensing, thermal imaging, and telecommunications \cite{Rogalski2009, Martyniuk2014, Gunapala2014, Zhong2023, Zhu2023a, Zhuge2017}. Initially, mesoscopic-scale inorganic semiconducting materials were the primary candidates for developing faster, smaller, and more sensitive detectors. However, recent efforts have shifted towards engineering inorganic, organic, and hybrid inorganic/organic semiconducting nanostructures to significantly enhance performance, reduce device size, and create flexible form factor devices for industrial applications \cite{Li2023, Ramakrishnan2023, Zhu2023a}. Promising approaches include designing III-V semiconductor nanowires, in situ-grown metal/semiconductor nanoparticles, and utilizing chemical vapor deposition or exfoliation techniques to produce 2D inorganic graphene/perovskite structures for optoelectronic devices \cite{Li2023, Wang2021, Bansal2023}.

During the past few decades, interest in the physics and applications of IR photodetectors with a tunable band gap for telecommunication wavelengths ($\sim 1.2-1.7~\mu$m) has grown, especially those based on III-V semiconductors such as InGaAs \cite{Brouckaert2007 , Unal2023}. Several InGaAs photodetector designs are commonly studied, such as PN or PIN junctions, avalanche, and metal-semiconductor-metal (MSM) photodetectors \cite{Unal2023}. MSM detectors are desirable for telecommunication applications due to their fast response times and ease of fabrication and integration with other electronics \cite{Soole1991, Kuhl1992, Salem1995, Wohlmuth1996, Aliberti2002, Kim2004}. 

A standard MSM photodetector consists of interdigitated metallic electrodes deposited on an active layer, such as InGaAs, forming two Schottky junctions in series. A finger electrode design is often chosen for simplicity of the device, both in terms of fabrication and characterization. MSM detectors can suffer from large dark currents due to relatively low Schottky barriers, however, this can be mitigated by introducing a lattice-matched InAlAs top layer to increase the barrier height. Performance improvements have been achieved through band gap tuning via In$_{x}$Ga$_{1-x}$As alloy concentration, Schottky barrier enhancement, and superlattice structures \cite{Smiri2020, Wohlmuth1996, Unal2023}. In this paper, we explore the use of plasmonic semiconductor nanocrystals to enhance the optical-to-electrical conversion efficiency of InGaAs MSM photodetectors, as measured by responsivity in the wavelength range of $1-1.7~ \mu$m.

Plasmonics primarily focuses on the generation, propagation, and detection of plasmonic waves, which are collective electronic excitations resulting from the interaction between an electromagnetic (EM) field and a metal/dielectric interface \cite{Berini2013, Dorodnyy2018, Zhang2022}. When EM waves interact with a metallic surface, a coherent charge density oscillation coupled to the EM field arises, known as a surface plasmon. Strong nanoscale confinement leads to localized surface plasmon resonances (LSPRs), characterized by discrete optical absorption bands and electric fields orders of magnitude larger than those associated with free space EM waves. Given their nanoscale dimensions, integration of plasmonic nanostructures with micro- and nanoelectronic devices is possible, and holds the potential to enhance device performance or realize novel functionalities.

Conductors with sub-wavelength dimensions support LSPRs, whose resonance frequencies depend on the geometry and free carrier density of the nanostructure, as well as its orientation with respect to the polarization of the EM field. Depending on the material, LSPRs can span from the ultraviolet (UV) to the far-infrared. Upon illumination, nanoscale regions of strong EM field enhancement, referred to as hot-spots, are generated at the surface of the structure. Plasmonics have been widely used to enhance the performance of photodetector and photovoltaic devices \cite{Butun2012, AzizurRahman2016, Goosney2019, Yang2012, Zhu2021, Yan2023, Lee2024}.  The enhanced EM field surrounding nanostructures can be harnessed in optoelectronic devices by generating LSPRs to improve responsivity across a broad range of frequencies. LSPR mechanisms that contribute to enhanced responsivity include: 1) increased concentration of the incident EM field and extended optical path length within the semiconductor, thereby enhancing photon absorption, and 2) plasmonic energy transfer from the nanostructure to the semiconductor through direct electron transfer \cite{Cushing2013}. Because of these broadband optical properties, plasmonic nanostructures have found use in a variety of applications including catalysis, medicine, surface chemistry, solar cells and sensing \cite{Berini2013, Dorodnyy2018, Zhang2022, Huang2018, Ji2019, Li2021, Alqanoo2022, Rajkumari2022, Zheng2023, Maity2023}.

Typically, Au and Ag nanoparticles (NPs) exhibit strong LSPRs and are used as plasmonic materials in the visible \cite{Liu2011} or UV \cite{Wu2018a, Tian2014, Kunwar2020} spectral ranges. An emerging area of interest focuses on identifying high performance plasmonic materials compatible with other frequency ranges of interest, from NIR to terahertz \cite{Senanayake2011, Lin2020, Zhu2023, Daugas2023}. Fundamentally, the physics of LSPR does not change in the IR compared to the visible region, but the material requirements are dramatically different. The optical resonances in Au and Ag can be tuned to a degree by changing the size and shape of nanostructures, but the charge carrier densities ($\sim$10$^{22}$ cm$^{-3}$) that primarily determine the the plasmonic frequency response are fixed. NIR applications require materials with lower carrier concentrations compared to metals, such as degenerately doped or non-stoichiometric semiconductors \cite{Nuetz1999, Wang2010, Luther2011, DellaGaspera2014}. There have been several reports on the use of materials such as transition metal oxides \cite{Dai2016, Fang2017}, nitrides \cite{Ahmadivand2016} and doped semiconductors \cite{Bao2014} for IR plasmonic photodetectors. 

Two candidate materials for IR plasmonics are WO$_{3-x}$ \cite{Zhu2023} and Cs$_y$WO$_{3-x}$ \cite{Daugas2023} nanocrystals (NCs). Compared to commonly used plasmonic Au/Ag nanostructures exhibiting LSPR in the UV-visible region, both WO$_{3-x}$ and Cs$_y$WO$_{3-x}$ possess visible-NIR LSPR due to their lower carrier densities, in the range of 10$^{18}$-10$^{22}$ cm$^{-3}$. These materials are known to have optical absorption in the NIR, but their effectiveness as plasmonic materials for enhancement of responsivity in a photodetector has not yet been thoroughly investigated; only one recent study has investigated photodetector enhancement in the visible light range using WO$_{3-x}$ nanowire arrays \cite{Zhu2023}. The carrier densities of these materials can also be tuned via the chemical synthesis procedure (e.g., extrinsic doping). However, a potential challenge when using WO$_{3-x}$ and Cs$_y$WO$_{3-x}$ NCs, despite being environmentally benign, is their rate of oxidation in air. Oxidation degrades the carrier concentration over time or with exposure to higher temperatures, which may lead to changes in crystal structure by reducing the oxygen vacancy concentration, and thus the carrier density.

In this paper, we characterize the performance of InGaAs MSM photodetectors operating in the NIR modified by drop-cast WO$_{3-x}$ nanorods and Cs$_y$WO$_{3-x}$ hexagonal prisms. Clear enhancements of responsivity by up to a factor $\sim$2.5, broadly consistent with the plasmonic absorption spectra of the NCs, are observed. These results indicate that suitably prepared conductive metal oxide NCs are a promising family of materials for IR plasmonic applications. 

\section{Materials and Methods}

\begin{figure}[b]
    \centering
    \includegraphics[width=\linewidth]{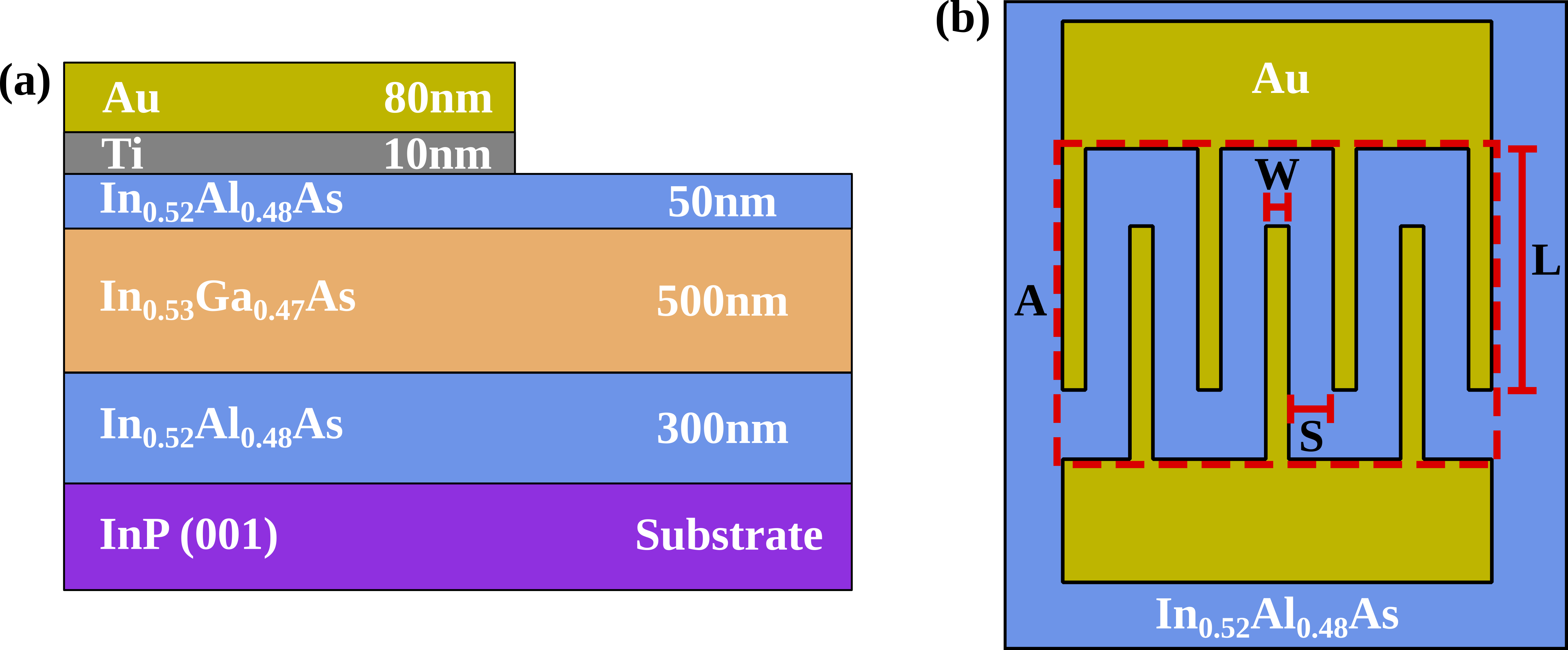} 
    \caption{Structure of the IR metal-semiconductor-metal (MSM) photodetectors. (a) Semiconductor heterostructure layers grown by molecular beam epitaxy. The In$_{0.53}$Ga$_{0.47}$As active region is grown between In$_{0.52}$Al$_{0.48}$As barriers on an InP substrate. (b) Top view schematic of the interdigitated metal electrode geometry, where design parameters are electrode width (W $\sim 10~\mu$m), spacing (S $\sim$ few~$\mu$m), length (L $\sim 500 ~\mu$m), and active area (A $\sim 500 ~\mu$m $\times$ 540$ ~\mu$m) indicated by dashed red box.}
    \label{fig:device_geometry}
\end{figure}

\subsection{Molecular Beam Epitaxy of InGaAs Heterostructures}

The InGaAs heterostructure shown in figure~\ref{fig:device_geometry}(a) was grown by molecular beam epitaxy (MBE). Prior to growth, a 2", semi-insulating (Fe-doped), double-side polished InP(001) substrate was heated to 520$~^\circ$C for approximately 10 minutes to desorb surface oxides. Epitaxy of the InAlAs/InGaAs/InAlAs layers was carried out at a substrate temperature of 470$~\pm 10~ ^\circ$C. Both the In$_{0.53}$Ga$_{0.47}$As and In$_{0.52}$Al$_{0.48}$As layers were grown at a rate of 2.5 \AA/s.

\subsection{Nanocrystal Synthesis}
\subsubsection{Materials}

 All chemicals were used as received, with no further purification. W(CO)$_6$ (97 \%) and HPLC-grade tetrachloroethylene (TCE) ($\geq$ 99.5 \%) were purchased from Thermo Scientific, WCl$_4$ (95 \%), CsCl (95 \%) from Kodak, technical-grade oleic acid (OlAc, 90 \%) and oleylamine (OlAm, 70 \%), HPLC-grade acetone ($\geq$ 99.9 \%), toluene ($\geq$ 99.9 \%), and hexane ($\geq$ 98.5 \%) were all purchased from Sigma Aldrich.

\subsubsection{Synthesis of WO$_{3-x}$}

Substoichiometric tungsten oxide (WO$_{3-x}$) NCs were synthesized following the thermal decomposition synthesis procedure for tungsten NPs reported by Sahoo et al. \cite{Sahoo2009}, with minor modifications as previously reported \cite{KennyWilby2023}. Briefly, 7.626 g of OlAc, 2.415 g of OlAm, and 1.057 g of W(CO)$_{6}$ precursor were mixed inside a reaction vessel and degassed for 30 minutes. Using a standard Schlenk line technique, the vessel was purged with nitrogen gas, then the reaction mixture was heated to 300 $^{\circ}$C at a rate of 0.5 $^{\circ}$C/s. After 2 hours at 300 $^{\circ}$C, the vessel was removed from the heating mantle and naturally cooled to room temperature. The blue-color reaction mixture, indicating the presence of WO$_{3-x}$ NCs, was purified three times using toluene and ethanol in 1:1 volumetric ratio. After each washing cycle, the samples were centrifuged at 3000 RPM for 5 min at room temperature. After final purification, the WO$_{3-x}$ NCs were dispersed in 5 mL of toluene. For the optical extinction measurements, 1 mL of the toluene dispersion was precipitated once again using the above method and redispersed in 1 mL of TCE. All the colloidal dispersions were stored under N$_{2}$ atmosphere until further use.

\subsubsection{Synthesis of Cs$_{y}$WO$_{3-x}$}

Hexagonal Cs$_y$WO$_{3-x}$ NCs were synthesized using the synthesis method reported by Mattox et al. \cite{Mattox2014} with minor modifications. The solid precursors (0.20 mmol of WCl$_4$ and 0.12 mmol of CsCl) were added into a mixture of OlAc (8.95 g) and OlAm (0.267 g), which was then degassed for approximately 20 min at 120 $^{\circ}$C. The reaction mixture was then heated to 300 $^{\circ}$C and kept at that temperature. After 2 hours, the reaction mixture was cooled down to room temperature, and the dark blue product obtained was pipetted into centrifuge tubes. NCs were washed with toluene and acetone in approximately 1:1 volumetric ratio applying vortex mixing for 5 minutes, then the sample was centrifuged at 7830 RPM for 5 min to separate the NCs from the supernatant. Similar to the WO$_{3-x}$ NCs, the as-prepared Cs$_y$WO$_{3-x}$ NCs were suspended in toluene or TCE for optical measurements and stored under N$_2$ atmosphere until further use.

\subsection{Photodetector Fabrication}

A 2 cm $\times$ 2 cm die from the as-grown InGaAs heterostructure wafer was cleaned by sonication in acetone and isopropyl alcohol (IPA) for 2 minutes each. Polydimethylglutarimide polymer (PMGI) and Shipley 1813 positive photoresists were deposited by spin coating at 5000 RPM for 60 seconds to create a homogeneous bilayer approximately 1.3 $\mu$m thick. The resist is baked at 120$^\circ$C for 90 seconds, followed by optical lithography performed using a Heidelberg instruments MLA150 maskless aligner to pattern the interdigitated electrodes and bond pads. The patterned resist was developed in a metal free optical developer (MF-319) for 100 seconds followed by a 20 second plasma ash to remove traces of resist residue from the exposed areas. A 1 minute HF etch was performed to remove surface oxides prior to deposition of 20/80nm of Ti/Au via electron beam evaporation. Metal liftoff was performed by leaving the sample overnight in Remover PG followed by rinsing in acetone and IPA. A schematic of the interdigitated electrode geometry is shown in figure~\ref{fig:device_geometry}(b). 

\subsection{Characterization of Nanocrystals}

X-ray diffractometry (XRD) was performed to study the crystallinity and phase of the synthesized NCs using a Rigaku MiniFlex II Diffractometer equipped with a Cu K$\alpha$1 radiation source ($\lambda$ = 1.54056 \AA). Transmission electron microscopy (TEM) measurements were performed to investigate the NC morphology using a JEOL-2010F instrument operating at 200 kV. The optical extinction spectra of the colloidal NC dispersions in TCE or NC thin films were performed to characterize the resonant wavelengths, absorption peak widths, and overall spectral shape using a Varian Cary 5000 UV-vis-NIR spectrophotometer. The synthesis and optical properties of the WO$_{3-x}$ and Cs$_y$WO$_{3-x}$ NCs in this work were found to be highly reproducible (see Figures S4-S5 in Supplementary Information).

\subsection{Photocurrent Measurement Setup}

\begin{figure}[h!]
    \centering
    \includegraphics[width=0.75\linewidth]{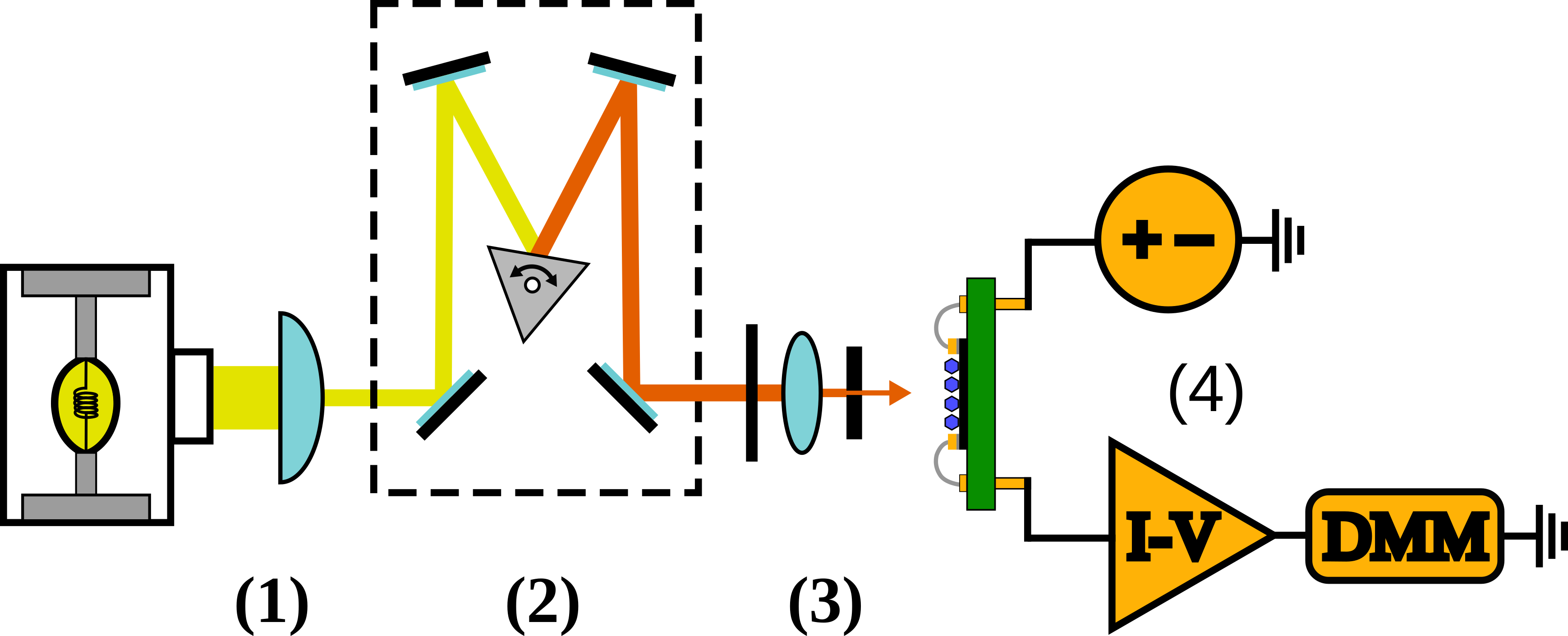} 
    \caption{Measurement setup for characterizing photodetector responsivity. (1) The broadband light source generates a beam that is collimated and then focused on the monochromator entrance slit; (2) wavelengths are selected by mechanical positioning of diffraction gratings; (3) the monochromator output is filtered and focused onto the wire-bonded MSM device; (4) a bias voltage is applied to the device and the photocurrent is measured using a current-voltage (I-V) amplifier. The output of the I-V amplifier is read out with a digital multimeter (DMM). Responsivity is determined from the DMM output and incident beam power measurements.}
    \label{fig:measurement_setup}
\end{figure}

Photodetectors are characterized using a custom setup that enables photocurrent measurements as a function of bias voltage and incident wavelength of light from the visible to the near-IR. The setup also enables characterization of the power spectrum of the light source, which is necessary for proper normalization of the measured photodetector responsivity. A schematic of the setup is shown in figure~\ref{fig:measurement_setup}. 

A Thorlabs SLS402 Mercury-Xenon short-arc light source generates a broadband spectrum ranging from 240 nm-2.4 $\mu$m that is collimated and focused using a Thorlabs LA4464-f = 60.0 mm UV fused plano-convex uncoated lens into the 2 mm aperture entrance slit of a Sciencetech monochromator. A desired wavelength is selected by choosing one of the three Blaze gratings inside the monochromator and rotating the grating into to proper position with respect to the incident beam. The wavelength ranges for each of the three gratings are 100-700 nm, 700-1500 nm, and 1500-3500 nm. The diffracted beam for a selected wavelength then passes through a 2 mm aperture exit slit, FEL1400 or FEL0750 longpass filters to prevent higher-order diffracted wavelength modes, and is then focused using a Thorlabs LA5370-f = 40.0 nm CaF$_2$ plano-convex uncoated lens through a post-mounted iris diaphragm onto the sample focal plane.  

Photodetector devices are attached to a printed circuit board and wire bonded to Au plated pins that connect to an external circuit through a custom socket. The sample mount is fixed to a lateral (XY) translation stage assembled from Thorlabs single-axis PT1 and PT series components to facilitate alignment of a chosen detector with the beam. A Keithley 230 programmable voltage source supplies a bias voltage to the source/drain electrodes of a detector. Current is measured using a DL instruments 1212 current-voltage amplifier, with the output signal fed into an Agilent 34401A digital multimeter. The monochromator and the voltage source are computer-controlled using a custom written LabView program. 

\subsubsection{Responsivity Measurement Protocol}

The broadband light source is switched on for 1 hour to reach a stable operating temperature and ensure stability of the output spectrum. Prior to measuring the photodetector spectral responsivity, a Thorlabs 700-1800 nm, 50 nW-40 mW Ge photodiode power sensor is used to obtain the power spectrum of the light source. The power sensor is placed in the focal plane behind the post-mounted iris diaphragm and the FEL0750 filter is placed in between the monochromator and LA5370 focusing lens. The power spectrum is measured over the wavelength range 750-1400 nm, with data points acquired at 5 nm intervals. In the 1400-1800 nm range, the FEL1400 filter is used. 

Once the source power spectrum is acquired, the power sensor is replaced by a chosen photodetector, which is then positioned using the XY translation stage to maximize the photocurrent detected at a reference wavelength of 550 nm. The photodetector dark current is recorded over a bias voltage range of +1 V to -1 V with the iris closed. The photocurrent versus bias voltage is then measured across the same wavelength values as the power spectrum to obtain the spectral responsivity of the device. All such measurements were carried out after covering the optical table with a piece of nylon blackout fabric to shield the detector from ambient light. 

\subsection{Nanocrystal Drop-Casts} \label{drop_cast_method}

Nanocrystals are deposited onto the active area of the photodetectors by drop-casting, using diluted suspensions of the WO$_{3-x}$ or Cs$_y$WO$_{3-x}$ NCs. Each colloidal NC suspension is diluted with hexane using a VWR single-channel pipette. Approximately 0.255 g of the original 190 g/L suspension of Cs$_y$WO$_{3-x}$ NCs was measured on a digital scale, and then diluted with 340 $\mu$L of hexane yielding a concentration of 152 g/L. Similarly, a WO$_{3-x}$ colloidal NC suspension was diluted to yield a concentration of 100 g/L. The dilute suspensions are then drop-cast onto the active areas of photodetectors using a 0.5 $\mu$L volume, and subsequently dried with nitrogen gas.

\section{Results}

\subsection{Nanocrystal Characterization}

\begin{figure}[b]
    \centering
    \includegraphics[width=\textwidth]{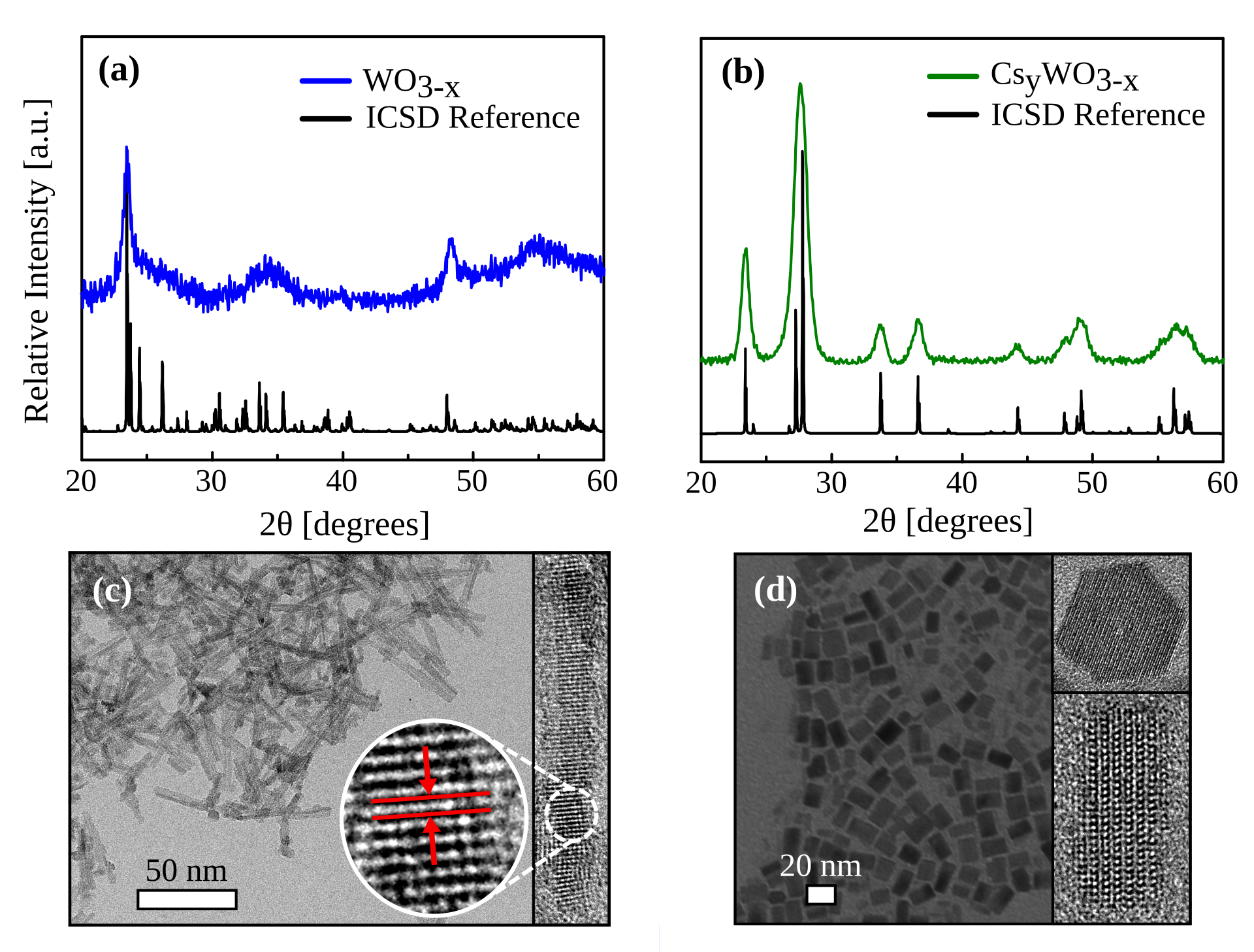}
    \caption{Characterization of crystalline structure and morphology of the synthesized NCs. (a-b) X-ray diffraction (XRD) spectra of WO$_{3-x}$ and Cs$_y$WO$_{3-x}$ NCs with background signal subtracted. Reference spectra of the bulk materials are reproduced from ICSD no. 24731 and no. 56223. (c) TEM images of drop-cast WO$_{3-x}$ nanorods; (right inset) lateral cross-section of a single nanorod; (magnified inset) lattice spacing along $\langle010\rangle$ direction. (d) TEM images of drop-cast Cs$_y$WO$_{3-x}$ NCs; (top right inset) end cross-section of a single NC; (bottom right inset) lateral cross-section image of a single NC.}
    \label{fig:NC_morphology}
\end{figure}

\begin{figure}[h!]
    \centering
    \includegraphics[width=\linewidth]{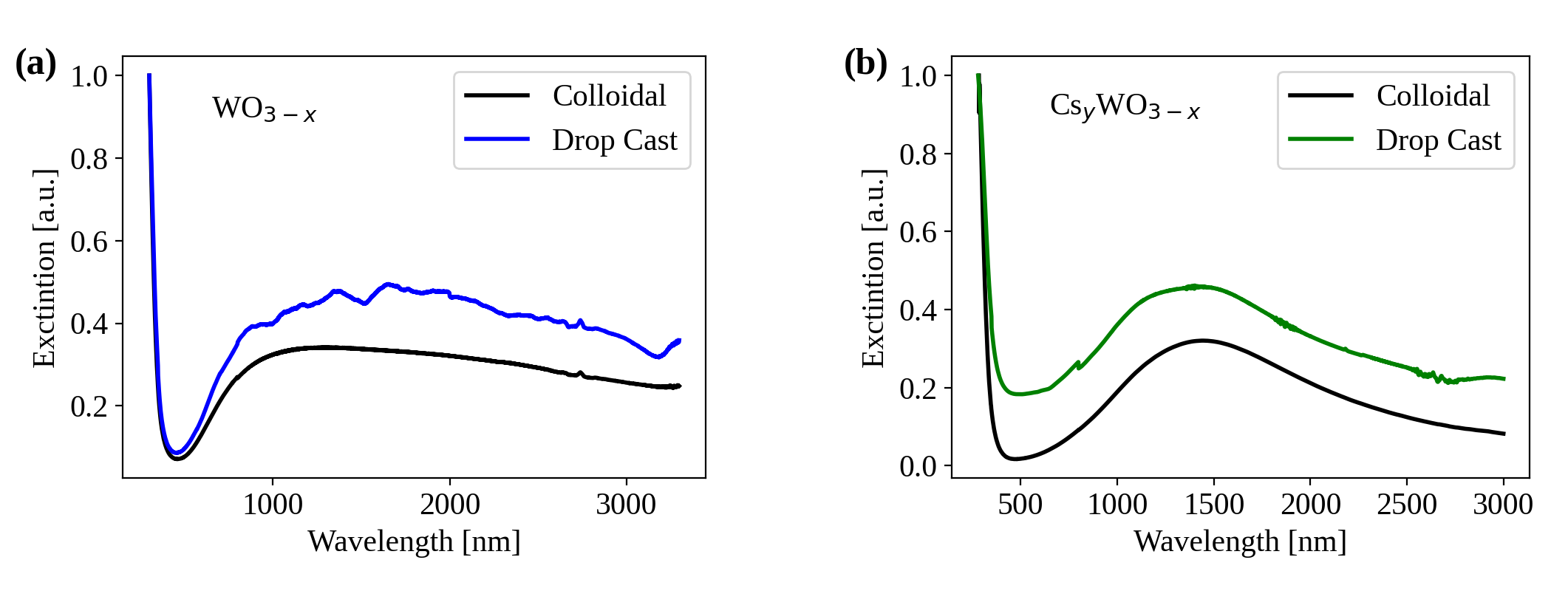}
        \caption{(a-b) Optical extinction spectra measured for colloidal solutions and drop-cast WO$_{3-x}$/Cs$_y$WO$_{3-x}$ NCs showing plasmonic resonance in the NIR-IR wavelength range. Colloidal solutions were prepared using tetrachloroethylene (TCE). Spectra are normalized by the peak extinction and vertically shifted for comparison.}
    \label{fig:extinction}
\end{figure}

Depending on the specific W/O atomic ratio, substoichiometric tungsten (VI) oxide (WO$_{3-x}$) can adopt different crystal structures, known as Magneli phases \cite{Migas2010}. In the present study, colloidally synthezised WO$_{3-x}$ NCs crystallized in the monoclinic phase, as suggested by the excellent match of the experimental X-ray diffractogram (figure \ref{fig:NC_morphology}(b), blue trace) with the simulated XRD pattern (black trace) of bulk monoclinic WO$_{2.72}$. The two dominant peaks in the XRD pattern of the WO$_{3-x}$ NCs appearing at approximately 23.5$^{\circ}$ and 48.3$^{\circ}$ can be assigned to (010) and (020) planes, respectively, suggesting the oriented growth of the NCs along the $\langle010\rangle$ direction. This observation was confirmed by the TEM micrographs seen in figure \ref{fig:NC_morphology}(c) showing the presence of elongated NCs (nanorods) with a mean length of 32.1 $\pm$ 7.8~nm and diameter of 4.3 $\pm$ 1.4~nm (for further details see Supplementary Information). The inset of figure \ref{fig:NC_morphology}(c) displays a high-resolution TEM micrograph of a WO$_{3-x}$ nanorod having a mean lattice spacing of 3.7~\r{A} along the $\langle010\rangle$ direction indicated by the red lines/arrows. In undoped non-stoichiometric plasmonic metal oxides, oxygen vacancies can introduce at most two electrons per vacancy ($\text{V}_{o}^{\bullet \bullet}$) into the crystal in order to maintain charge neutrality \cite{Lounis2014}, which can subsequently lead to the reduction of nearby metal ions in the lattice (\textit{e.g.}, $2\text{W}^{6+} +  \text{V}_{o}^{\bullet \bullet} \rightarrow 2 \text{W}^{5+} + \text{V}_{o}$). To investigate the electronic structure and tungsten speciation in the WO$_{3-x}$ nanorods, X-ray photoelectron spectroscopy (XPS) measurements were performed, the results of which were presented in our earlier report \cite{KennyWilby2023}. The W~4f XPS spectra suggested approximately 6\% W$^{5+}$ to be present in the WO$_{3-x}$ nanorods. In the optical extinction spectrum of the colloidal (black trace) and drop-cast (blue trace) WO$_{3-x}$ NCs shown in figure \ref{fig:extinction}(a), a characteristic broad band starting in the visible region near 490 nm and extending into the NIR region can be observed. Based on the earlier report \cite{KennyWilby2023}, the optical absorption in the visible region originates from intra-ionic (d-d) transitions, whereas the absorption at lower energies in the NIR region are assigned to LSPR originating from the collective oscillation of oxygen-vacancy-induced free electrons in the conduction band. The intense absorption in the UV region with an onset starting near 480~nm is assigned to the excitonic (band-to-band) transition of the semiconducting WO$_{3-x}$ nanorods \cite{KennyWilby2023}.

When tungsten oxide is doped with metal ions, a family of non-stoichiometric compounds of the general formula M$_y$WO$_{3}$ (M is usually an alkali metal ion) is formed, called tungsten bronzes \cite{Conroy1952}. Depending on the type and radius of the dopant ion, the crystal structure as well as the free carrier concentration, and thus the metallic conductivity, can be modulated in tungsten bronzes. The X-ray diffractogram seen in figure \ref{fig:NC_morphology}(b) reveals that as-synthesized Cs$_y$WO$_{3-x}$ NCs are of high purity and crystallized in the hexagonal phase, indicated by the excellent match between the experimental data and the simulated XRD pattern of the hexagonal-phase Cs$_{0.29}$WO$_{3}$. The two most intense peaks near 23.3$^{\circ}$ and 27.7$^{\circ}$ originate from the (002) prismatic and (200) basal planes, respectively. The TEM images shown in figure \ref{fig:NC_morphology}(d) reveal a rod-like morphology for as-synthesized Cs$_y$WO$_{3-x}$ NCs and a mean length and height of 17.8 nm and 10.4 nm (aspect ratio 1.7), respectively (see figure S2 of the Supplementary Information). The high-resolution TEM micrographs in figure \ref{fig:NC_morphology}(d) display an inset image of a basal plane (top) and the prismatic plane (bottom) of a Cs$_y$WO$_{3-x}$ NC along the $\langle002\rangle$ direction. In tungsten bronzes the electric charge introduced by the dopant ion results in an increase in the number of free electrons in the conduction band leading to an increase in electrical conductivity. Therefore, besides the oxygen vacancies, the other source of free carriers in Cs$_y$WO$_{3-x}$ NCs is the cesium dopant ion. In the optical extinction spectra shown in figure \ref{fig:extinction}(b), the sharply increasing UV signal originates from the bandgap absorption, similar to WO$_{3-x}$ NCs. The broad vis-NIR absorption band of Cs$_y$WO$_{3-x}$ nanorods has been reported in the literature to originate from two resonance (transverse and longitudinal) modes as a result of strong crystalline and shape anisotropy \cite{Cheref2022, Kim2016}, with an additional polaronic band in the visible region \cite{Machida2019}. Given the geometric parameters of the as-synthesized Cs$_y$WO$_{3-x}$ NCs (aspect ratio 1.7), the longitudinal and transverse LSPR modes nearly overlap, resulting in the broadened single-peak absorption band seen in figure 4(b) \cite{Cheref2022, Kim2016}.

\subsection{Photodetector Characterization}
\begin{figure}[htb!]
    \centering
    \includegraphics[width=0.9\textwidth]{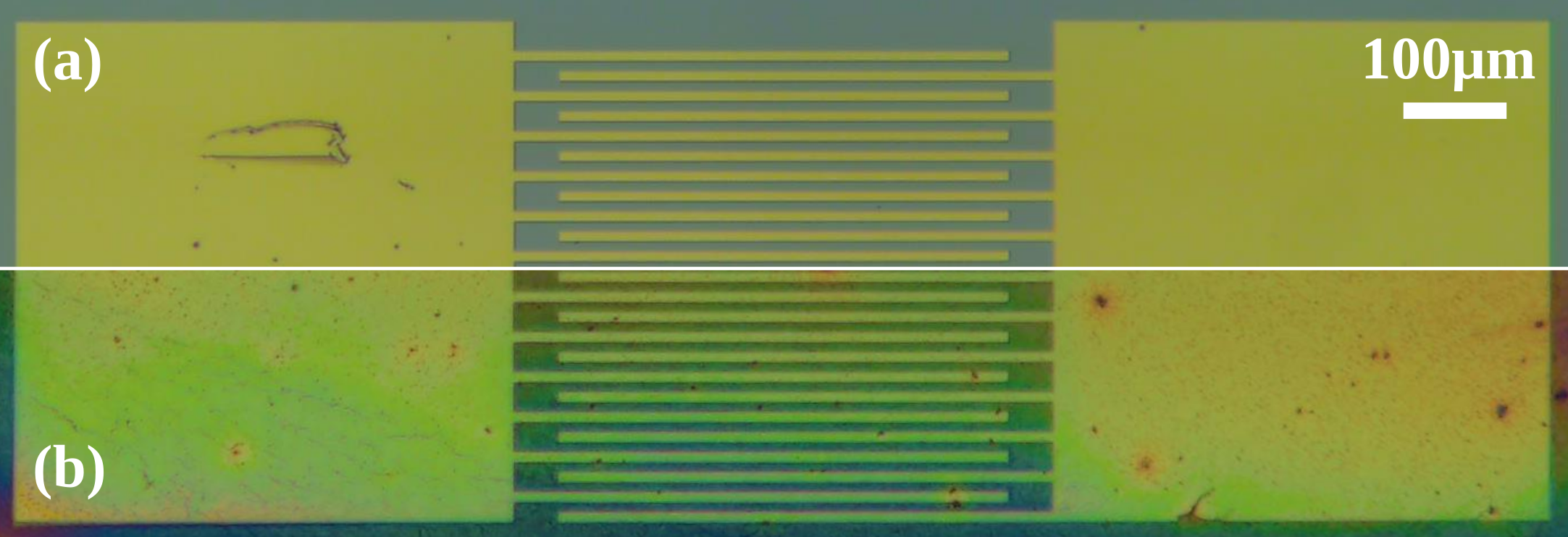}
    \caption{Optical microscope images of a MSM photodetector with a similar geometry to the photodetectors measured in this study: a) Upper half of the MSM device with no NCs b) lower half of the MSM device after drop-casting Cs$_y$WO$_{3-x}$ NCs from a 150 g/L solution. The lower image indicates a relatively uniform distribution of NCs.}
    \label{fig:msm_dropcast}
\end{figure}

An array of 16 MSM photodetectors with varying electrode geometries was characterized using an Agilent E5262A source/measure unit probe station. The typical device resistances were measured under white light illumination ($R_{light}$) and in the dark ($R_{dark}$). Figure \ref{fig:msm_dropcast}(a) shows an optical microscope image of a photodetector device with geometry similar to those used in this study. The interdigitated finger electrode structure allows colloidally suspended NCs to be readily drop-cast, as discussed previously in section \ref{drop_cast_method}. 

\noindent Table \ref{tab:device_resistances} summarizes the resistance values measured with applied bias -10 mV to +10 mV, including the mean and standard deviation values for the entire array. Two devices, referred to as A and B, showed relatively large increases in conductivity under illumination, and were selected for further measurements.

\begin{table}[htb!]
    \centering
    \begin{tabular}{|c|c|c|c|}
    \hline
    Device & $R_{\text{dark}} [M\Omega]$ & $R_{\text{light}} [M\Omega]$ & $\Delta R [\%]$ \\
    \hline
    $\mu \pm \sigma$ All & 4.24 $\pm$ 2.44 & 0.144 $\pm$ 0.0950 & -96.6 \\
    A & 0.200 & 0.0183 & -90.8 \\
    B & 4.55 & 0.181 & -96.0 \\
    \hline
    \end{tabular}
        \caption{Resistance values for as-fabricated photodetectors measured on a probe station with ($R_{light}$) and without ($R_{dark}$) illumination under a white light source to characterize photodetector performance. The applied bias ranged from -10 mV to +10 mV. The optical power was approximately 18~$\frac{\mu W}{mm^2}$ with the probe station lamp on and 32~$\frac{nW}{mm^2}$ with the lamp turned off (ambient light). Devices A and B were chosen for the plasmonic enhancement study because they exhibited significant decreases in resistance under illumination and had the same electrode geometry. $\Delta R$ refers to the percentage change in resistance under illumination. The mean resistance ($\mu$) and standard deviation ($\sigma$) across the array of 16 detectors are also indicated. \label{tab:device_resistances}}
\end{table}

\begin{figure}[h!]
    \centering
    \includegraphics[width=\textwidth]{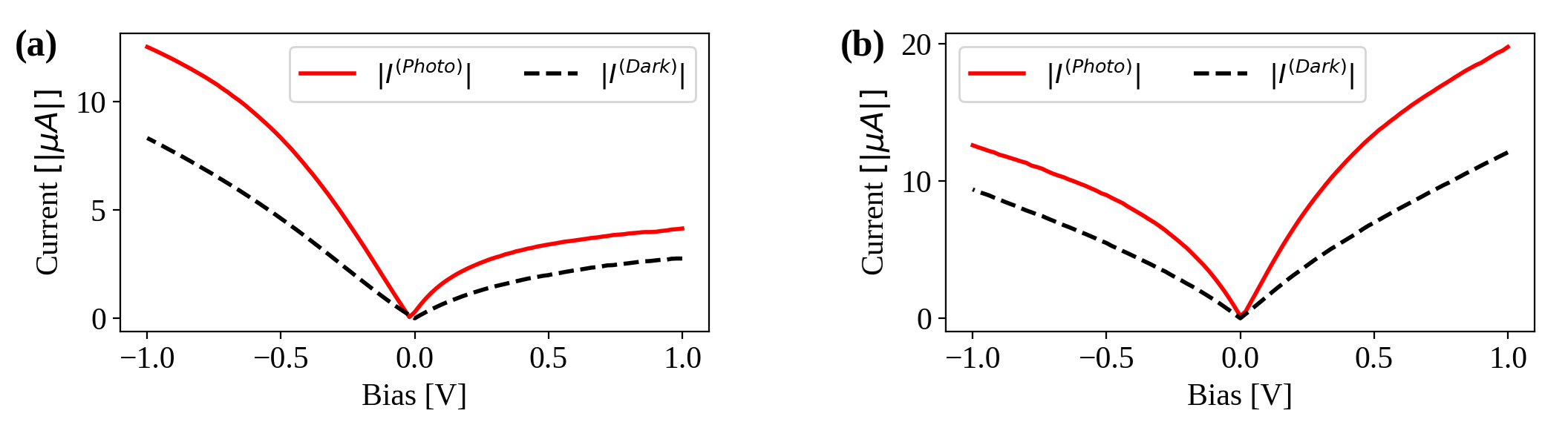}
    \caption{Characterization of photodetector performance versus bias voltage and illumination. Dark current and photocurrent for the as-fabricated InGaAs photodetectors A (a) and B (b) used in this study. Photocurrent was acquired at a wavelength of 1550 nm with an incident beam power of 176 $\frac{\mu W}{mm^2}$ over the photodetector active area.}
    \label{fig:msm_device_ide6}
\end{figure}

\noindent The spectral responsivities of devices A and B were measured using the setup described in figure~\ref{fig:measurement_setup}. These photodetectors share the same geometry, with active area 500~$\mu$m $\times$ 540~$\mu$m, and finger width, spacing and length of 10~$\mu$m, 3~$\mu$m, and 500~$\mu$m, respectively. The photoresponse versus bias voltage obtained at wavelength of 1550 nm is shown in figure~\ref{fig:msm_device_ide6}. A relatively large dark current was observed compared to similar InGaAs MSM photodetectors reported in the literature ($\sim$nA \cite{Wohlmuth1996, Soole1991}), even with the InAlAs barrier layer present. 
We attribute this to a pitting surface defect density of $\sim$$10^6~\text{cm}^{-2}$ in the InAlAs/InGaAs wafer material that lead to regions with lowered Schottky barriers and deviations from typical thermionic emission behavior \cite{Breitenstein2006, Chistokhin2019}. Using a double Schottky barrier model described in \cite{Bhattacharya2022}, the dark current through the MSM device can be modeled as
\begin{align}
    J(V) &= \frac{2 J_{S}(\Phi_{1}) J_{S}(\Phi_{2}) \sinh \left(\frac{\beta V }{2n} \right)}{J_{S}(\Phi_{1}) \exp \left( \frac{\beta V }{2n} \right) + J_{S}(\Phi_{2}) \exp \left( \frac{-\beta V }{2n} \right)}
\end{align} \label{eq: dark_current}
\noindent where $\beta = 1/(k_B T)$, $\Phi_{1/2}$ is the Schottky barrier for either contact, $n$ is the ideality factor assumed to be the same for both contacts, and $J_{S}(\Phi_{1/2})$ are the thermionic saturation current densities for either contact. From the double Schottky barrier model, the average barrier heights between MSM contacts were estimated to be $\langle \Phi_A \rangle \approx$ 0.53~eV and $\langle \Phi_B \rangle \approx$ 0.50~eV for both devices A and B, compared to the theoretically expected 0.7~eV \cite{Chistokhin2019}. These results are also consistent with earlier theoretical predictions \cite{Wohlmuth1996}. See the Supplementary Information for characterization of wafer morphology and details of the dark current model.

\subsection{Enhancement of Photodetector Spectral Responsivity}

\begin{figure}[b]
    \centering
    \includegraphics[width=\textwidth]{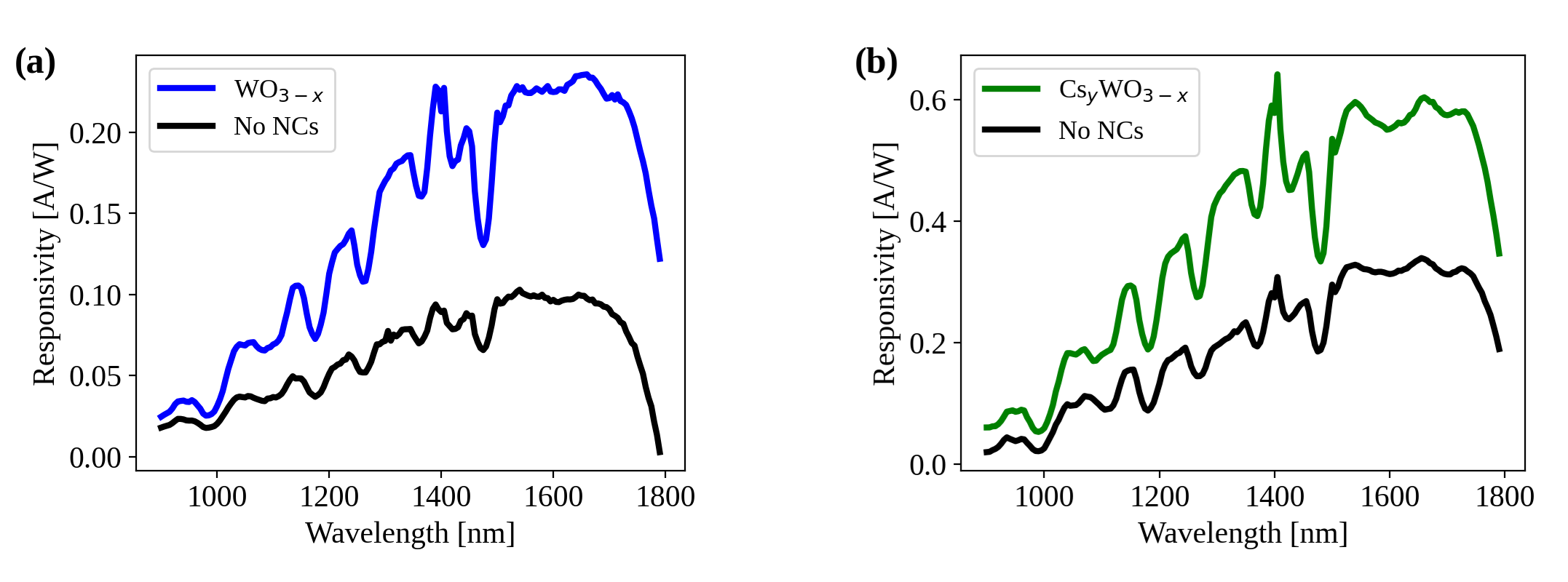}
    \caption{Responsivity measured at a bias voltage -1 V with and without (a) WO$_{3-x}$ NCs (device A) and (b) Cs$_{y}$WO$_{3-x}$ NCs (device B). Significant increases in responsivity are observed for both types of NC.  Responsivity is determined by subtracting the dark current and normalizing the photocurrent with respect to the measured power spectrum of the light source. Note that the fluctuations in responsivity versus wavelength are reproducible and not due to measurement noise (uncertainty due to noise is smaller than the plot line width). These fluctuations may arise from sharp features in the broadband light source spectrum that are not fully removed by our normalization protocol (see Supplementary Information) or by wave interference due to the electrode geometry.}
    \label{fig:response}
\end{figure}

Spectral responsivity is defined as the ratio of the output photocurrent to the incident optical power, and it measures the optical-to-electrical conversion efficiency of the photodetector. The spectral responsivities of the as-fabricated photodetectors were first determined by measuring the dark current, photocurrent and power spectrum of the light source. A fixed bias voltage of -1 V was used for both devices. Note that the light source power spectrum was appropriately scaled to account for the differing areas between the photodiode detector and the MSM photodetectors, as well as variations in the beam profile. Also, measurement artifacts from switching between two gratings during the spectral sweep were removed (see Supplementary Information). 
Next, diluted solutions of WO$_{3-x}$ (100 g/L) and Cs$_y$WO$_{3-x}$ (152 g/L) NCs were drop-cast onto the active area of each photodetector and dried. The optical microscope image in figure~\ref{fig:msm_dropcast}(b) shows an example of drop-cast Cs$_y$WO$_{3-x}$ NCs. Similar drop-casts are obtained for WO$_{3-x}$ NCs, although there can be significant variability in the quality of the drop-cast (see Supplementary Information). The responsivity measurements were then repeated in order to quantify the effect of having NCs present.

Figure \ref{fig:response} shows the responsivities with and without NCs for devices A and B. Device A (B) had WO$_{3-x}$ (Cs$_y$WO$_{3-x}$) NCs drop-cast. Before NC deposition, both devices show a wavelength dependent responsivity characteristic of MSM photodetectors with an InGaAs active layer \cite{Chen2022, Li2022}. Using the responsivity, the specific detectivity for both devices is calculated as:
\begin{align}
    D^* = \sqrt{\frac{A}{2 q_e I_{dark}}} \times R(\lambda)
\end{align} \label{eq: detectivity}

\noindent Where $A$ is the electrode contact area, $q_e$ is the elementary electron charge, $I_{dark}$ is the average dark current, and $R(\lambda)$ is the wavelength-dependent responsivity. Peak detectivities of $D^*_A\sim$$10^9$~Jones and $D^*_B\sim$$10^{10}$~Jones are observed for devices A and B, respectively, which are comparable to other NIR photodetectors \cite{Bansal2020, Bansal2023, Lee2024} that typically range from $D^*\sim$$10^8-10^{14}$~Jones. Furthermore, MSM InAlAs/InGaAs photodetector dectivities up to $D^*\sim$$10^{11}$~Jones can be achieved with improved substrate quality \cite{Kim2004}. Significantly increased responsivity is observed for the addition of both WO$_{3-x}$ and Cs$_y$WO$_{3-x}$ NCs to the active regions of the photodetectors.

\subsubsection{Relative Enhancement}
It is of interest to examine the relative enhancement of responsivity as a function of wavelength and compare with the optical extinction measurements of the drop-cast NCs. The relative percent difference (RPD) is defined as:

\begin{equation} \label{eq:RPD}
    RPD(\lambda) = \frac{R_{NC}(\lambda) - R_{0}(\lambda)}{R_{0}(\lambda)} \times 100 \%
\end{equation}

\noindent where $\lambda$ is wavelength, and $R_{NC}(\lambda)$ and $R_{0}(\lambda)$ are the responsivities with and without NCs, respectively. 

\begin{figure}[h!]
    \centering
\includegraphics[width=\linewidth]{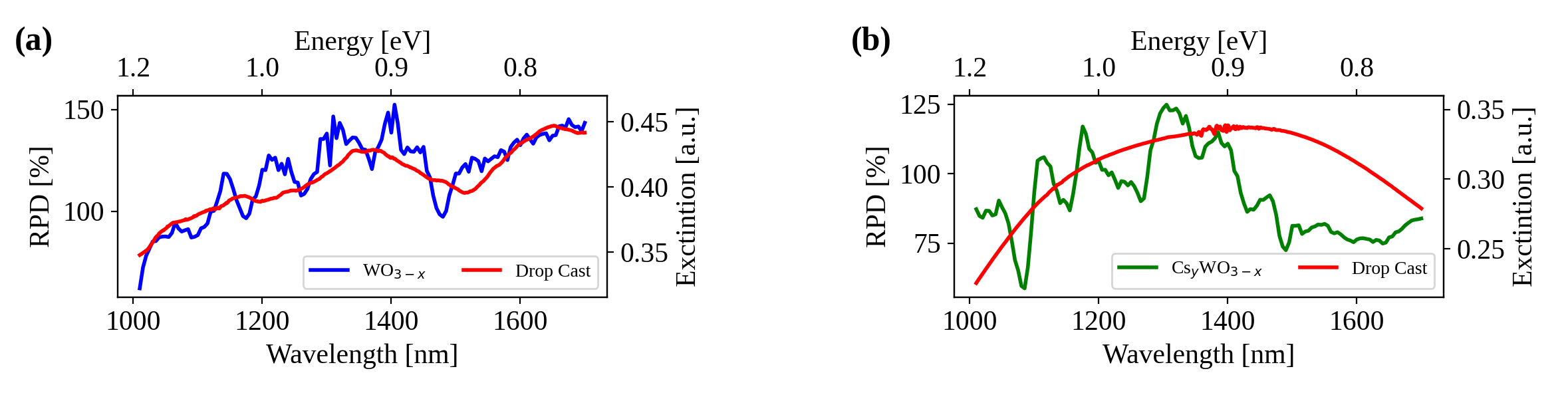}
    \caption{Relative percent change in responsivity (RPD) as a function of wavelength due to the addition of (a) WO$_{3-x}$ and (b) Cs$_y$WO$_{3-x}$ NCs. The optical extinction measurement for drop-cast NCs (red line), plotted on the right vertical axis in each plot, shows overall agreement with the RPD. Note that the relative scaling of the two vertical axes is arbitrary.}
    \label{fig:RPDvsExtinction}
\end{figure}

The RPD is plotted in figure \ref{fig:RPDvsExtinction}, with the NC optical extinction measurements overlaid. The RPD calculated from equation \ref{eq:RPD} is in the range of 60-150\% and 58-125\% for devices A and B, respectively, demonstrating a strong overall enhancement of the responsivity with the addition of both NC types. 
\noindent Broad agreement between the RPD and the NC extinction spectra is observed in figure \ref{fig:RPDvsExtinction}, indicating that the plasmonic responses of the NCs likely plays a role in the enhanced photoresponse of the MSM detectors. Better agreement between the RPD and the NC extinction spectra was observed for WO$_{3-x}$ NCs compared to the case of Cs$_y$WO$_{3-x}$ NCs. The details of the spectral responses are expected to depend strongly on aggregation effects and film thickness, both of which can vary significantly between drop-casts (see Supplementary Information). 

\subsubsection{External Quantum Efficiency}
\begin{figure}[h!]
    \centering
    \includegraphics[width=\linewidth]{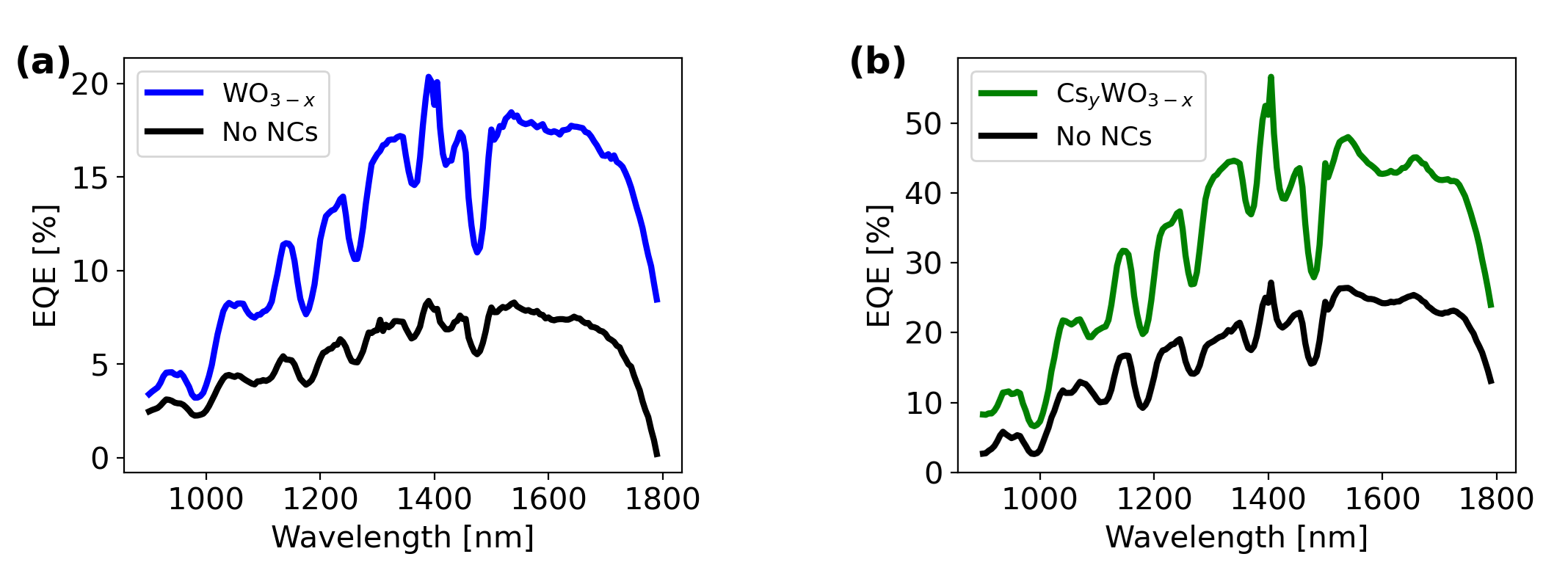}
    \caption{External quantum efficiency (EQE) of each photodetector with and without the addition of (a) WO$_{3-x}$ and (b) Cs$_y$WO$_{3-x}$ NCs. EQEs based on the data shown in figure~\ref{fig:response} demonstrate improved photodetector optical-to-electrical conversion efficiency for both NC types.}
    \label{fig:EQE_enhancment}
\end{figure}

\noindent The ability of a photodetector to convert incident photons to charges collected in the electrodes is quantified by the external quantum efficiency (EQE) as a function of wavelength \cite{Shi2020}:
\begin{equation} \label{eq:EQE}
    EQE(\lambda) = \frac{R(\lambda) h c}{q_e \lambda} \times 100 \%
\end{equation}
\noindent where $h$ is Planck's constant and $c$ is the speed of light. Figure \ref{fig:EQE_enhancment} shows the EQE for both photodetectors with and without NCs, calculated based on the data shown in figure~\ref{fig:response}. The highest EQEs are observed in the wavelength range 1400-1600 nm for both devices. 

The EQE increases from approximately 7\% to 20\% for device A and approximately 25\% to 60\% for device B at a wavelength of 1400 nm, indicating that $\sim$2-3 times as many charge carriers are generated or collected after the addition of the WO$_{3-x}$ and Cs$_y$WO$_{3-x}$ NCs. The EQEs before the addition of NCs are comparable to those reported for other NIR photodetectors \cite{Duan2023} and are enhanced by the addition of NCs.

\section{Discussion}

Mechanisms for plasmonic response giving rise to increased photodetector EQEs have been debated in recent years. Possible mechanisms to explain our observations include:  
 
\begin{enumerate}
    \item LSPR modes of the NCs \cite{Cheref2022, Kim2016} create stronger electric fields in the InGaAs layer of the MSM photodetector, generating charge carriers at a higher rate \cite{Abdulhalim2018}.
    
    \item Plasmon-induced hot carriers are injected from the NCs into the bulk semiconductor \cite{Tian2005, Chu2023, Clavero2014, Luo2023, Lee2024}.
    
    \item Photon collection from scattering is increased due to the NC films forming an anti-reflective (AR) coating \cite{Katagiri2014, Zeb2021, Cushing2013}.
    
\end{enumerate}

\noindent It is not clear from the present data which of these mechanisms are present or dominant in our devices. However, recent work has demonstrated that injection of plasmonically-excited  carriers can play a primary role in enhanced photodetection \cite{Lee2024, Duan2023}. Evidence that hot carrier injection is the primary mechanism for WO$_{3-x}$ nanorod photodetectors has been demonstrated \cite{Zhu2023}. The plasmonic absorption band for our NCs ranges over 0.7-1.2~eV. Depending on the work function of the NCs, a Schottky barrier could form that would be low enough to inject energetic carriers into the bulk semiconductor, generating a photocurrent under an applied bias. To test this hypothesis, carrier injection could be suppressed by depositing a thin, wide-bandgap insulating layer between the NCs and the InAlAs surface. The influence of hot carrier injection on photodetector responsivity could be studied by varying the doping concentration of the NCs, where a higher doping concentration would be expected to yield a stronger increase in responsivity. 
Additionally, LSPR-generated EM fields can be orders of magnitude stronger than the incident field. A stronger EM field in the InGaAs active layer would generate electron-hole pairs at a higher rate. Here, the enhancement in photodetection would be governed by the thickness of the InAlAs buffer layer, since the electric field generated at the NC surface drops off as $|E| \propto 1/r^2$ \cite{Abdulhalim2018}. This mechanism can be tested by varying the thickness of the InAlAs layer.

The magnitude and distribution of the EM field is determined by the NC morphology which governs the packing behavior of the NCs and the dominant LSPR mode \cite{Kim2016}. The optimal morphology for increased density of EM field hot spots in the active region could be investigated using finite-difference time-domain EM simulations \cite{Kim2016, Cheref2022}, and correlated with responsivity measurements for different NC morphologies.

Finally, enhancement in EQE from the NCs acting as an AR coating could be potentially challenging to separate from other LSPR effects. Plasmonic nanostructures have been used for AR coatings \cite{Singh2020}, alongside commonly used dielectric nanoparticles. A systematic study of the reflectivity of transition metal oxide NC films on various surfaces could clarify the significance of this effect.

Future work will aim to clarify the dominant mechanisms for enhancement and improve performance through the following approaches:

\begin{enumerate}
    \item Synthesizing Cs$_y$WO$_{3-x}$ NCs with varied aspect ratios and carrier concentrations to elucidate the role of LSPR;
    \item Characterizing performance improvements using non-plasmonic dielectric AR coatings;
    \item Improving sensitivity by using higher-quality substrates for MBE growth to reduce dark current. In addition, incorporating control devices with variable-thickness dielectric barriers between the NCs and the InAlAs surface to test the hot carrier injection hypothesis;
    \item Applying NCs via cold vacuum aerosol deposition, or spin coating with NCs embedded in a silica matrix, in order to improve the reproducibility and quality of NC coverage.
\end{enumerate}

\section{Conclusion}

In this study we have demonstrated that easy-to-fabricate MSM photodetectors can exhibit improved performance in the NIR with the addition of semiconducting WO$_{3-x}$ or Cs$_y$WO$_{3-x}$ NCs to the active area. This shows the potential of both non-stoichiometric and doped transition metal oxide plasmonic nanomaterials for various IR optoelectronics applications, such as enhanced speed and sensitivity of receivers in optical fiber communications or increased range and reliability of Light Detection and Ranging (LiDAR) systems for autonomous vehicles. The ease of deposition by drop-casting is convenient for rapid prototyping of modified MSM detectors, but lacks the consistency needed for industrial scaling.

In future work, the mechanisms for plasmonic enhancement of photodetection can be distinguished by systematic experiments, as previously discussed, along with comparison to finite-difference time-domain EM simulations. Other directions for improvement include the use of higher quality substrates for MBE growth (reduced dark current), the use of more sophisticated NC deposition methods for reproducibility and consistency, and the use of a transparent conductive material (such as indium tin oxide) for the electrodes. 

\section*{Acknowledgments}

This research was undertaken thanks in part to funding from the Canada First Research Excellence Fund (Transformative Quantum Technologies) and the Natural Sciences and Engineering Research Council (NSERC) of Canada. The University of Waterloo's QNFCF Facility was used for this work. This infrastructure would not be possible without the significant contributions of CFREF-TQT, CFI, ISED, the Ontario Ministry of Research and Innovation, and Mike and Ophelia Lazaridis. Their support is gratefully acknowledged.

\section*{Author contributions}

ZDM performed the experimental work characterizing photodetectors and NC solutions for drop casting, data analysis and plotting, and led the writing of the manuscript. GyJ synthesized and characterized NC samples and contributed to writing the manuscript.
AJ fabricated the photodetectors.
HW grew the InGaAs wafer.
JB and PVR supervised the project and contributed to writing the manuscript. All authors reviewed the manuscript.   

\section*{Data availability statement}

The data that support the findings of this study are available upon request from the authors. 

\section*{ORCID iDs}

Zach D. Merino https://orcid.org/0000-0001-6224-800X

\noindent Gyorgy Jaics https://orcid.org/0009-0007-6210-4736

\noindent Pavle V. Radovanovic https://orcid.org/0000-0002-4151-6746

\noindent Jonathan Baugh https://orcid.org/0000-0002-9300-7134

% supplemental ---------------------------------------------------------------------------
\newpage

\section{Supplementary Information}

\subsection{Supplement A: Nanocrystal (NC) Characterization}

\begin{figure}[H]
    \centering
    \includegraphics[width=0.8\textwidth]{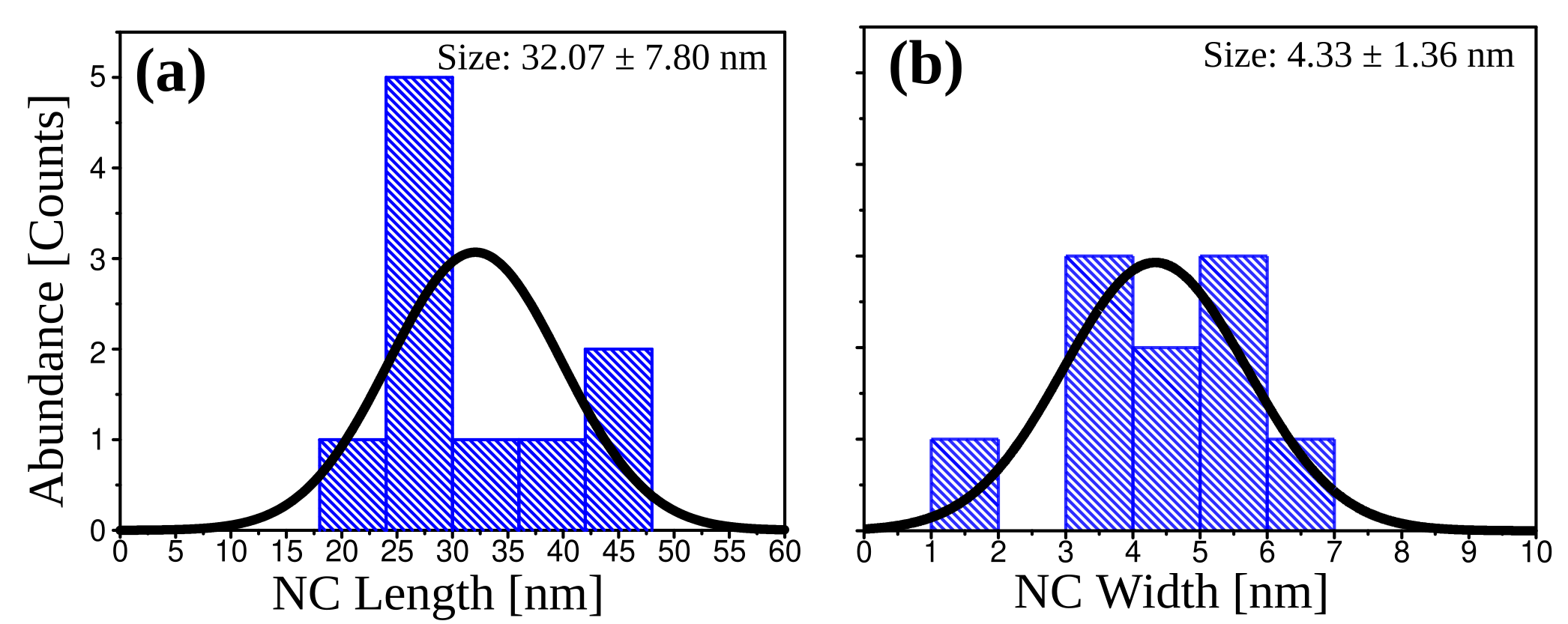}
    \caption{Length and width distributions of WO$_{3-x}$ NCs determined from post-processed TEM images using ImageJ software.}
    \label{supp-fig:wo_nc_sizes}
\end{figure}

\begin{figure}[H]
    \centering
    \includegraphics[width=0.8\textwidth]{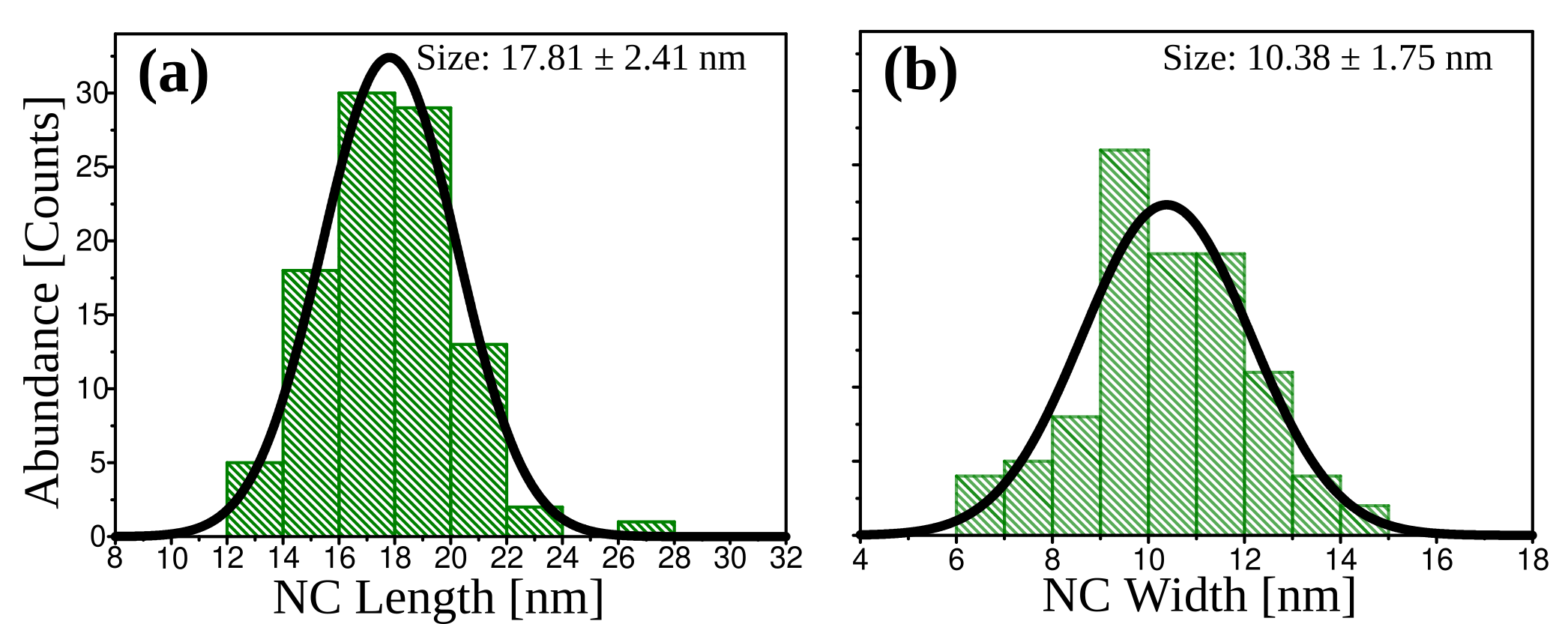}
    \caption{Length and width of distributions Cs$_y$WO$_{3-x}$ NCs determined from post-processed TEM images using ImageJ software.}
    \label{supp-fig:cswo_nc_sizes}
\end{figure}

\begin{figure}[H]
    \centering
    \includegraphics[width=0.75\textwidth]{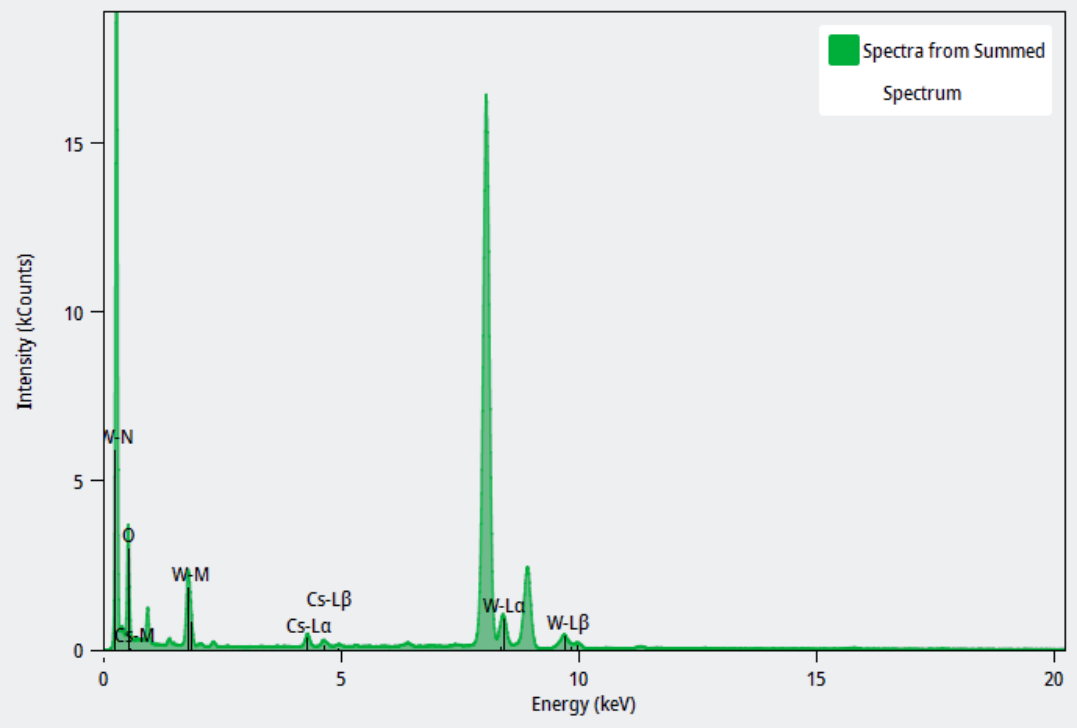}
    \caption{Energy dispersive X-ray spectroscopy (EDS) measurement on Cs$_y$WO$_{3-x}$ NCs that determined the atomic fraction of Cs and W to be 4.78\% and 17.67\%, respectively. From this, the ratio of Cs/W in Cs$_y$WO$_{3-x}$ NCs was determined to be $y=0.27$.}
    \label{supp-fig:cswo_eds}
\end{figure}

\begin{figure}[H]
    \centering
    \includegraphics[width=0.9\textwidth]{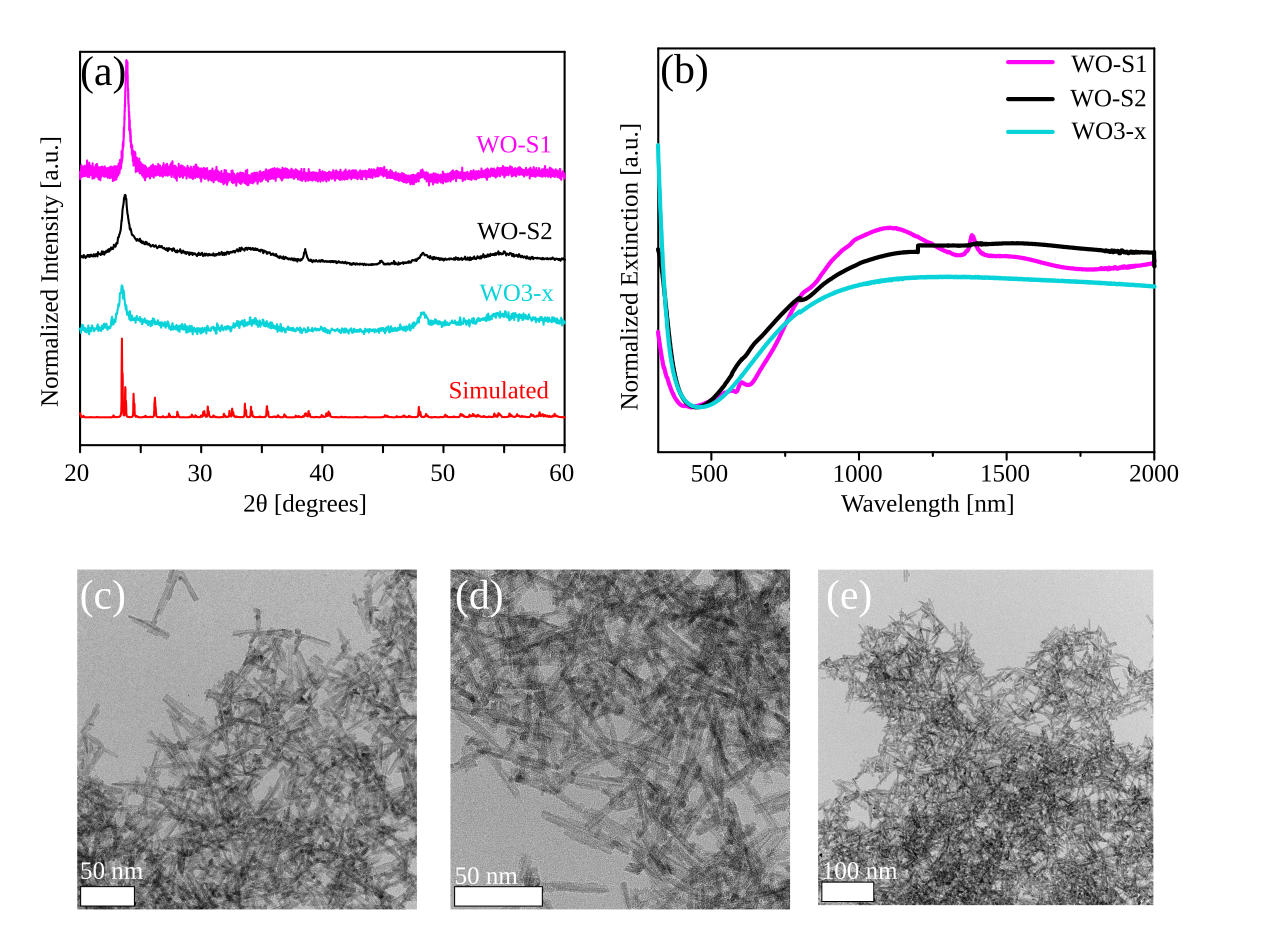}
    \caption{Characterization of two additional substoichiometric monoclinic WO$_{3-x}$ NC samples demonstrating the synthetic reproducibility of the NCs via XRD patterns (a), UV-Vis-NIR spectra (b), and (c-e) TEM micrographs. (Blue traces in (a) and (b) correspond to the WO$_{3-x}$ in main text. Red trace in (a) is simulated pattern of bulk monoclinic WO$_3$ (ICSD no. 24731).}
    \label{supp-fig:wo_samples}
\end{figure}

\begin{figure}[H]
    \centering
    \includegraphics[width=0.9\textwidth]{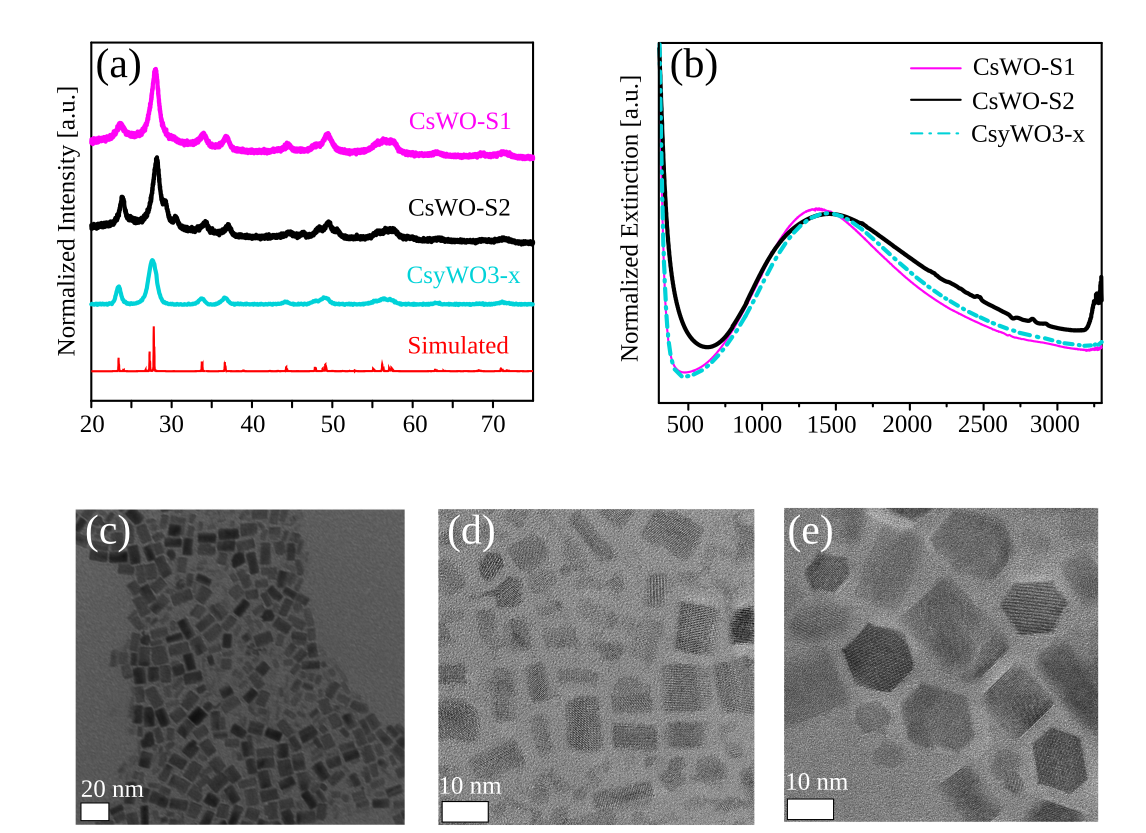}
    \caption{Characterization of two additional Cs$_y$WO$_{3-x}$ bronze NCs demonstrating the synthetic reproducibility of the samples. (a) XRD patterns, (b) UV-Vis-NIR spectra, and (c-e) TEM micrographs of the Cs$_y$WO$_{3-x}$ NCs. (Blue traces in (a) and (b) are the data shown in the main text. The red trace in (a) is the simulated pattern of bulk Cs$_{0.29}$WO$_3$ (ICSD no. 56223)).}
    \label{supp-fig:cswo_samples}
\end{figure}

Figures \ref{supp-fig:wo_nc_sizes} and \ref{supp-fig:cswo_nc_sizes} provide information on the size and shape distributions of the WO$_{3-x}$ and Cs$_y$WO$_{3-x}$ NCs, as determined from TEM images. Figure \ref{supp-fig:cswo_eds} shows energy dispersive X-ray spectroscopy results that were used to determine the composition of the Cs$_y$WO$_{3-x}$ NCs.  Figures \ref{supp-fig:wo_samples} and \ref{supp-fig:cswo_samples} show the crystalline and optical characterization for WO$_{3-x}$ and Cs$_y$WO$_{3-x}$ NCs, respectively. Tables \ref{supp-table:wo_table} and \ref{supp-table:cswo_table} present a summary of these results.

\begin{table}[H]
    \centering
    \caption{Summary of crystallographic and optical characterization of WO$_{3-x}$ NC samples from independent syntheses. The maximum plasmonic resonant wavelengths from extinction measurements (UV-NIR) and peak XRD angles for a given crystal orientation with Full-Width Half-Maximum ((crystal orientation)-FWHM) are tabulated. This demonstrates a high degree of reproducibility in the synthesis of WO$_{3-x}$ NCs.} \label{supp-table:wo_table}
    \begin{tabular}{|>{\centering\arraybackslash}p{2cm}|>{\centering\arraybackslash}p{1.75cm}|>{\centering\arraybackslash}p{2.1cm}|>{\centering\arraybackslash}p{2.1cm}|}
        \hline
        \textbf{Sample} & \textbf{UV-NIR} \newline Max [nm] & \textbf{XRD} \newline (010) & \textbf{XRD} \newline (020) \\
        \hline
        WO3-x & 1126  & 23.4-0.5 & 48.3-0.8\\
        \hline
        WO-S1 & 1098  & 23.8-0.5 & 48.3-0.6 \\
        \hline
        WO-S2 & 1153  & 23.7-0.7 & 48.3-1.3 \\
        \hline
    \end{tabular}
\end{table} 

\begin{table}[H]
    \centering
    \caption{Summary of crystallographic and optical characterization of Cs$_y$WO$_{3-x}$ NC samples from independent syntheses. The aspect ratios (AR), maximum plasmonic resonant wavelengths from extinction measurements (UV-NIR), and peak XRD angles for a given crystal orientation with Full-Width Half-Maximum ((crystal orientation)-FWHM) are tabulated. This demonstrates a high degree of reproducibility in the synthesis of Cs$_y$WO$_{3-x}$ NCs.}\label{supp-table:cswo_table}
    \begin{tabular}{|>{\centering\arraybackslash}p{1.75cm}|>{\centering\arraybackslash}p{0.75cm}|>{\centering\arraybackslash}p{1.75cm}|>{\centering\arraybackslash}p{1.25cm}|>{\centering\arraybackslash}p{1.25cm}|>{\centering\arraybackslash}p{1.25cm}|>{\centering\arraybackslash}p{1.25cm}|}
        \hline
        \textbf{Sample} & \textbf{AR} & \textbf{UV-NIR} \newline Max [nm] & \textbf{XRD} \newline (002) & \textbf{XRD} \newline (200) & \textbf{XRD} \newline (112) & \textbf{XRD} \newline (202) \\
        \hline
        CsyWO3-x & 1.6 & 1442 & 23.4-0.8 & 27.7-1.3 & 33.8-0.8 & 36.7-0.9 \\
        \hline
        CsWO-S1 & 1.5 & 1363 & 23.6-0.8 & 28.0-1.4 & 34.2-0.9 & 37.0-0.8 \\
        \hline
        CsWO-S2 & 1.6 & 1449 & 23.8-0.8 & 28.1-1.3 & 33.9-1.0 & 36.9-0.9 \\
        \hline
    \end{tabular}
\end{table}

\subsection{Supplement B: Light Source Characterization}
\begin{figure}[H]
    \centering
    \includegraphics[width=\linewidth]{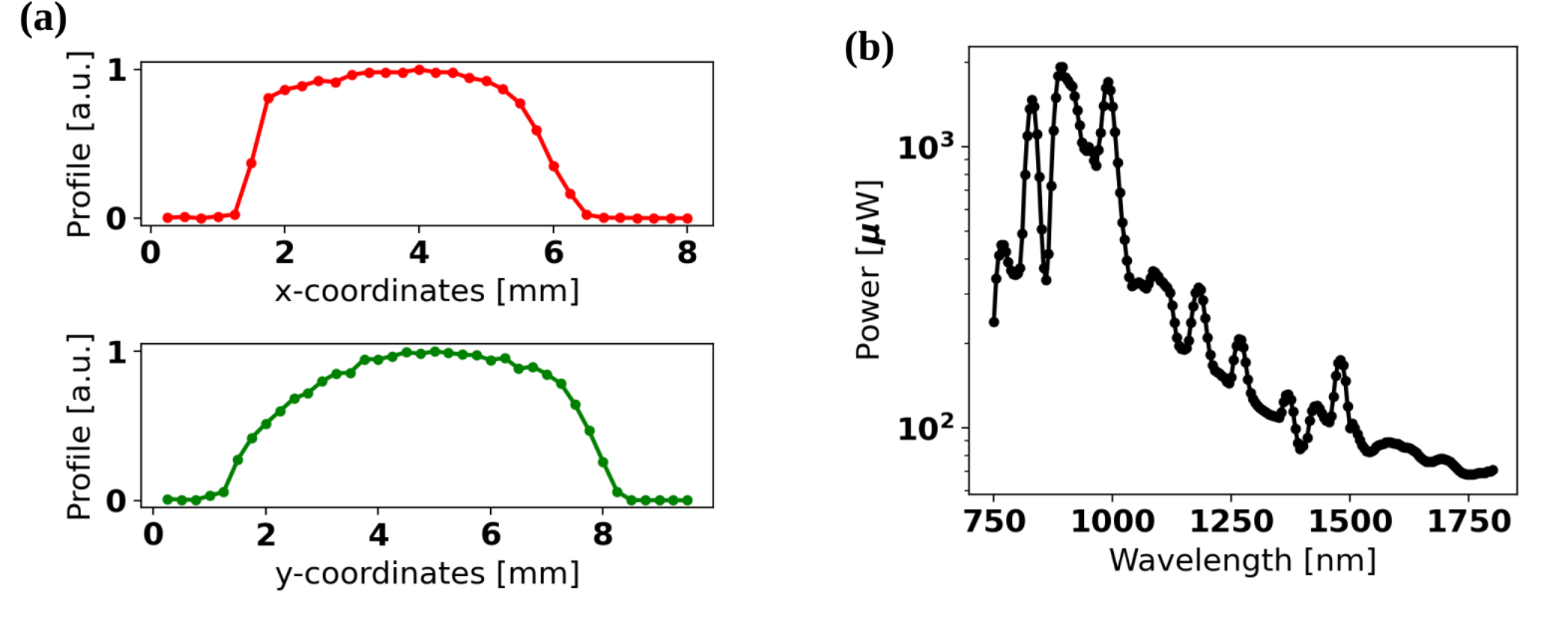}
    \caption{Xenon-Halogen light source (a) beam profile upon exiting the monochromator and (b) 
    power spectrum over a wavelength range 750-1800 nm where a 750 nm or 1400 nm filter was used to block higher order modes of diffracted light from the monochromator gratings.}
\end{figure}

Cross-sections of the beam profile are shown attached to a head on image of the beam at 550 nm wavelength light. The image of the beam is not to scale with the cross-sectional beam profiles. The beam profiles were measured using a standard knife-edge measurement.

\subsection{Supplement C: Spectral Responsivity Normalization and Incident Power}
\begin{figure}[H]
    \centering
    \includegraphics[width=\linewidth]{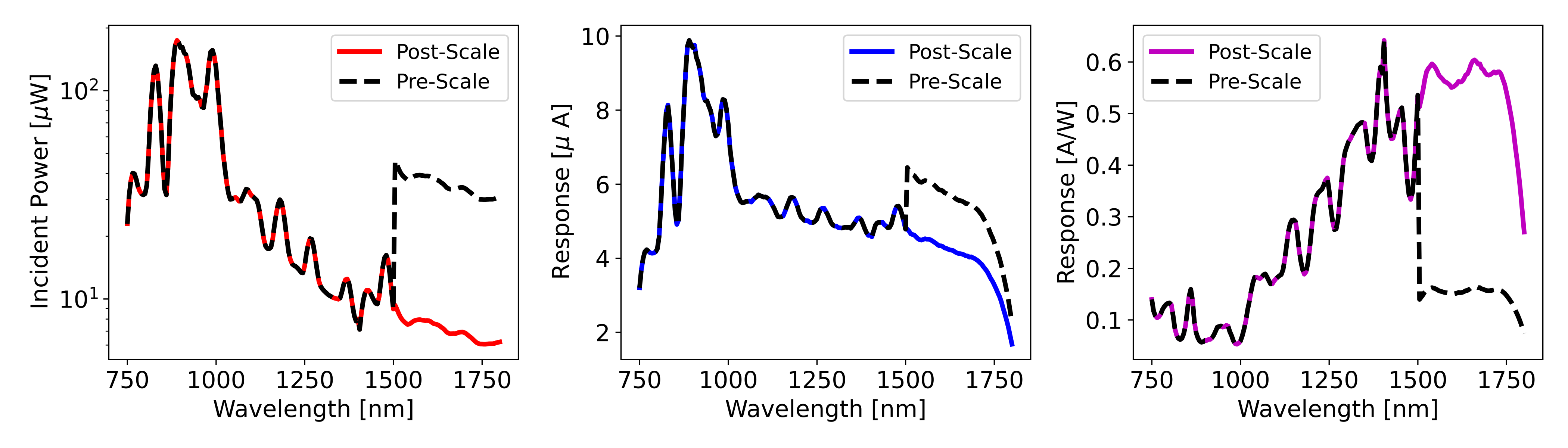}
    \caption{Example showing how the power spectrum and photocurrent data are processed to account for changes in power transmission through different diffraction gratings used in the monochromator. Two gratings are used over the range of wavelength values selected; the grating change occurs at 1500 nm. \label{supp-fig:scaling_spectra}}
\end{figure}
The total integrated power, $P_{total}(\lambda)$, incident on a power sensor photodiode area is recorded over a band of wavelengths. From the beam profile measurements above, the photodetector was positioned in a region of the beam profile with maximal power density, $\Delta P/\Delta r$. For the typical photodetector active area, the power density is nearly constant across the device. Calculating the total power incident on the photodetector, $P_{device}(\lambda)$, is performed by numerical integration of $\Delta P/\Delta r$ along $\Delta r \in \{\Delta x,\Delta y\}$ cross sections over the dimensions of the device active area position relative to the beam profile. The $P_{device}(\lambda=550\text{nm})$ was calculated from a beam profile measurement taken at a wavelength of 550 nm. Assuming the beam profile does not vary with wavelength, $P_{device}(\lambda)$ was calculated as the normalized total integrated power:
\begin{equation} \label{eq:power}
    P_{device}(\lambda) \coloneqq \frac{P_{device}(\lambda=550\text{nm})}{P_{total}(\lambda=550\text{nm})} \times P_{total}(\lambda)
\end{equation}
This method is a reasonable lower bound approximation for the incident power seen by the photodetectors. Since refractive optics were used for beam collimation and focusing, the beam spot size will slightly increase for $\lambda > 550 \text{nm}$ due to chromatic aberrations, decreasing the power density. 

Measuring the photodetector current or the light source power spectrum yields discontinuities in the measured spectra due to the change of diffraction grating at a wavelength of 1500 nm. The power density at the focal plane of the photodetectors changes due to power transmission variation between the two diffraction gratings used to perform each wavelength sweep measurement. The discontinuity is removed by properly normalizing the current and power spectra for the wavelength range associated with one of the diffraction gratings. The wavelength range which is normalized is 1500-1800 nm and the spectra are normalized by $\alpha$ or $\beta$ defined as:
\begin{equation} \label{eq:scale_factor}
    \alpha, \beta \coloneqq \frac{f^{(I, P)}_{G1}(\lambda=1500\text{nm})}{f^{(I, P)}_{G2}(\lambda=1500\text{nm})}
\end{equation}

\noindent where G1/G2 refer to diffraction blaze gratings (G1 $\coloneqq$ grating 1, G2 $\coloneqq$ grating 2) utilized for $f^{(I, P)}_{G1}(\lambda=1500\text{nm})$ and $f^{(I, P)}_{G2}(\lambda=1500\text{nm})$ which represent the measured quantities: (current $I \coloneqq f^{(I)}$) or (power $P \coloneqq f^{(P)}$). The normalized response is then calculating using the scale factors $\alpha$ and $\beta$:
\begin{equation} \label{eq:response}
    R(\lambda) = \frac{I_{\lambda_1} \cup \left( \alpha \times I_{\lambda_2} \right)}{P_{\lambda_1} \cup \left( \beta \times P_{\lambda_2} \right)}
\end{equation}

\noindent where $\lambda_1 \in [750~\text{nm}, 1500~\text{nm} ]$, $\lambda_2 \in (1500~\text{nm}, 1800~\text{nm} ]$, and $\cup$ represents the union of two data sets collected from measurements.

The power and photocurrent measurements are scaled independently such that the last/first values in the wavelength ranges 750-1500 nm and 1505-1800 nm are equivalent for both power and photocurrent spectra. The pre- and post-scaled spectra are plotted in figure \ref{supp-fig:scaling_spectra} where the dashed lines are the pre-scaled portions of the respective spectra.

\newpage

\subsection{Supplement D: In$_{0.52}$Al$_{0.48}$As/In$_{0.53}$Ga$_{0.47}$As Wafer Morphology}

As stated in the main text, we observed larger than expected dark currents in the MSM photodetectors, despite having an In$_{0.52}$Al$_{0.48}$As barrier layer present. From Nomarski microscope images (see figure S6) of the InAlAs surface, pitting defects can be observed with a density of $\sim$10$^6$cm$^{-2}$. Such defects can provide low resistance pathways between the metal electrodes and InGaAs active layer. Based on previous MBE growths in the same system, we suspect that these defects originate from defects native to the InP growth wafer. 

\begin{figure}[H]
    \centering
    \includegraphics[width=\linewidth]{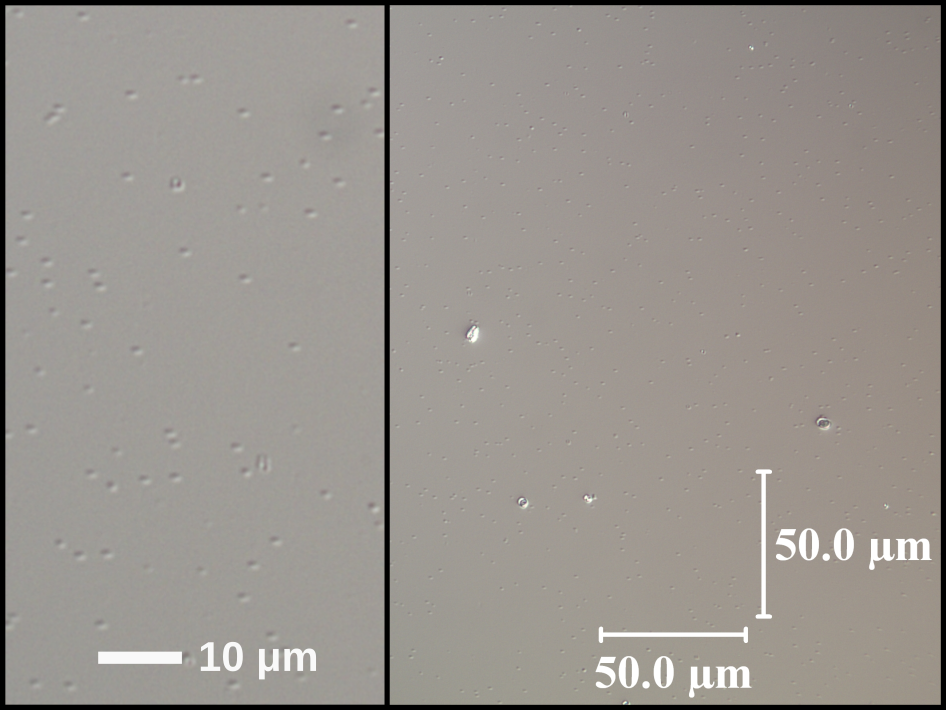}
    \caption{Characterization of surface defects using a Nomarski microscopy scan of MBE-grown In$_{0.52}$Al$_{0.48}$As/In$_{0.53}$Ga$_{0.47}$As on InP, taken at the center of the wafer. (left inset) Magnified view of pitting surface defects suspected to result from defects in the InP growth substrate.}
    \label{supp-fig:G0997_wafer}
\end{figure}

\subsection{Supplement E: Schottky Contact Model}

\begin{figure}[H]
    \centering
    \includegraphics[width=\linewidth]{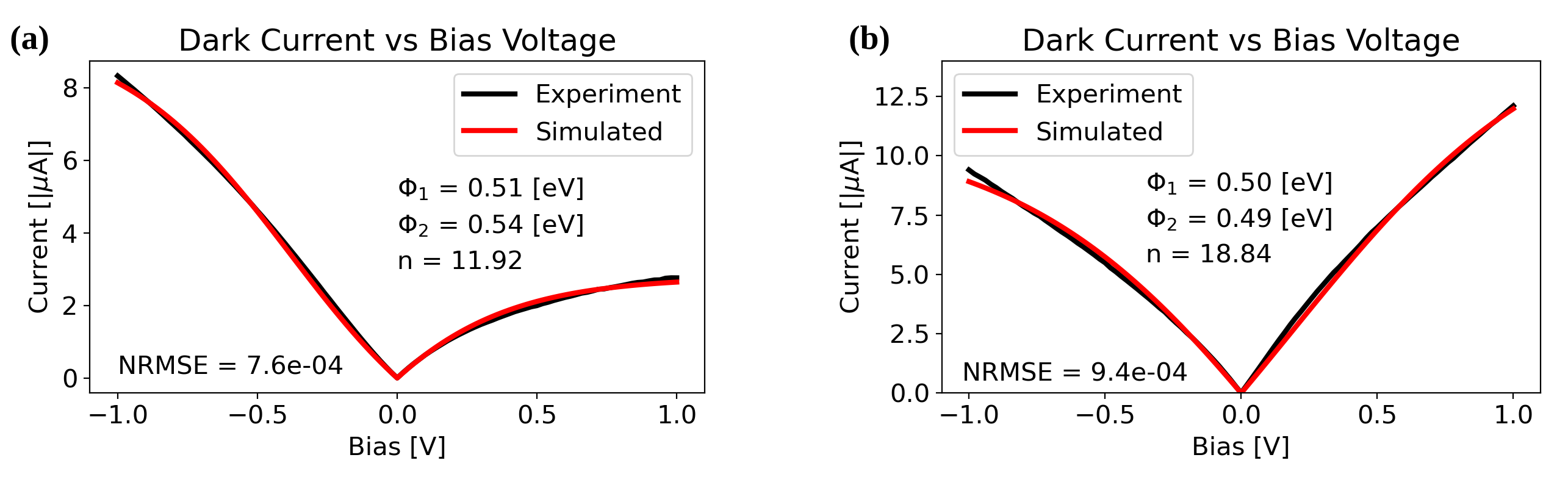}
    \caption{Dark current versus bias voltage for the as-fabricated InGaAs photodetectors A (a) and B (b) used in this study (black lines). Plotted in red are simulated I-V curves from the double Schottky barrier model described above.  Schottky barriers ($\Phi$), ideality factors (n), and Normalized Root Mean Square Errors (NRMSE) determine by the fitting are listed.}
    \label{supp-fig:msm_fit}
\end{figure}

Using a thermonic emission model for a double Schottky diode from \cite{Bhattacharya2022}, the dark current through the device is given by:
\begin{align*}
    J(V) &= \frac{2 J_{S}(\Phi_1) J_{S}(\Phi_{2}) \sinh \left(\frac{\beta V }{2n} \right)}{J_{S}(\Phi_{1}) \exp \left( \frac{\beta V }{2n} \right) + J_{S}(\Phi_{2}) \exp \left( \frac{-\beta V }{2n} \right)},
\end{align*}
\noindent where
\begin{align*}
    J_{S}(\Phi_{1/2}) &= A_{1/2} A^* T^2 \exp \left( - \beta \Phi_{1/2}\right),
\end{align*}

\noindent and where $\beta = 1/k_B T$, $\Phi_{1/2}$ is the Schottky barrier for either contact, $n$ is the ideality factor assumed to be the same for both contacts, $A_{1}= A_{2} = 3.342 \times 10^{-7} m^{-2}$ is the area of the contact, $A^*= 1.01 \times 10^5$ A m$^{-2}$ K$^{-2}$ is the Richardson constant for InAlAs, and $J_{S}(\Phi_{1/2})$ are the thermionic saturation current densities for either contact. From fitting our dark current data to this model, the Schottky barriers and ideality factor are estimated to be: (device A) $\Phi_{1} = 0.51$~eV, $\Phi_{2} = 0.54$~eV, and $n = 11.92$ (device B) $\Phi_{1} = 0.50$~eV, $\Phi_{2} = 0.49$~eV, and $n = 18.84$. The theoretically expected Shottky barrier for Ti/InAlAs is 0.7 eV \cite{Bhattacharya2022}. Figure \ref{supp-fig:msm_fit} shows each model fit and the experimental dark current data with the Normalized Root Mean Square Error (NRMSE) calculated using the equation NRMSE $=$ RMSE$/ [max(I_{dark}) - min(I_{dark})])$. NRMSE$_A = 7.6 \times 10^{-4}$ and NRMSE$_B = 9.4 \times 10^{-4}$ for devices A and B, respectively.

\subsection{Supplement E: Spectral Specific Detectivity $D^*(\lambda)$}

\begin{figure}[H]
    \centering
    \includegraphics[width=\linewidth]{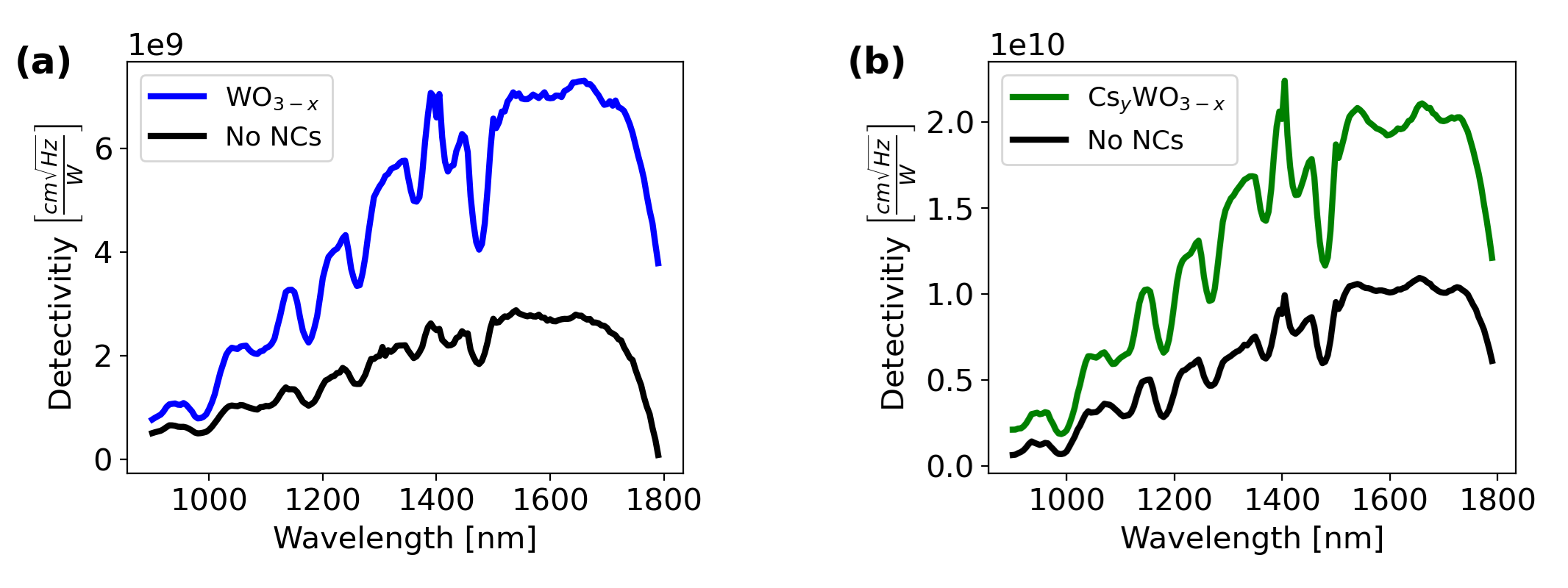}
    \caption{Specific Detectivity (in Jones, i.e. $\frac{cm \sqrt{Hz}}{W}$) is plotted as a function of incident wavelength for devices A and B, with and without the addition of NCs. Maxmum detectivity prior to adding NCs is $\sim$10$^9$ Jones for device A and $\sim$10$^{10}$ Jones for device B.}
    \label{supp-fig:detectivity}
\end{figure}

Figure \ref{supp-fig:detectivity} plots the Specific Detectivity as a function of wavelength for both photodetectors used in this study, before and after NCs were deposited in the active areas.

% \newpage

\subsection{Supplement F: Photodetector Drop-Casts}

\begin{figure}[H]
    \centering
    \includegraphics[width=0.85\linewidth]{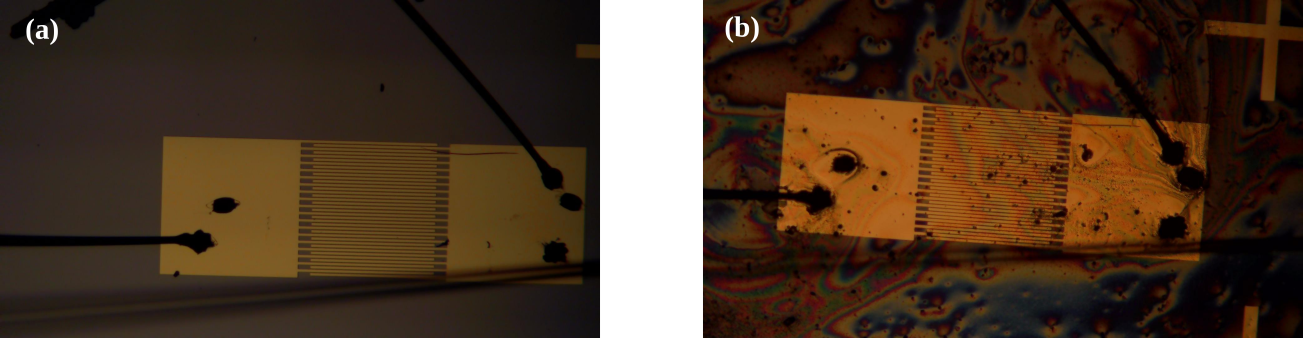}
    \caption{Optical image of a photodetector taken at 5X magnification (a) before and (b) after drop-casting of WO$_{3-x}$ NCs. The NC concentration and quality of the drop-cast yielded an increased photoresponse compared to the bare photodetector.}\label{supp-fig:dropcast_worked}
\end{figure}

\begin{figure}[H]
    \centering
    \includegraphics[width=0.85\linewidth]{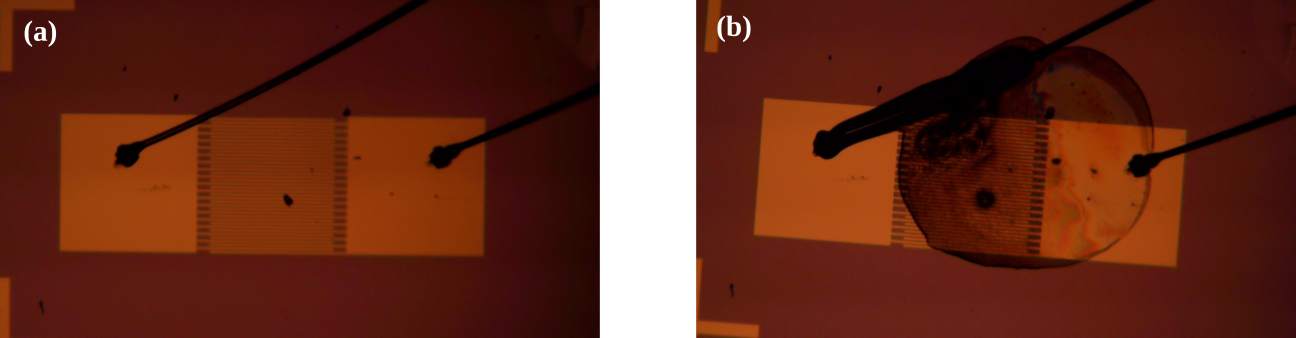}
    \caption{Optical image of a photodetector taken at 5X magnification (a) before and (b) after drop-casting Cs$_y$WO$_{3-x}$ NCs. The NC concentration and poor quality of the drop-cast yielded a decreased photoresponse compared to the bare photodetector.}\label{supp-fig:dropcast_not_worked}
\end{figure}

\begin{figure}[H]
    \centering
    \includegraphics[width=0.9\linewidth]{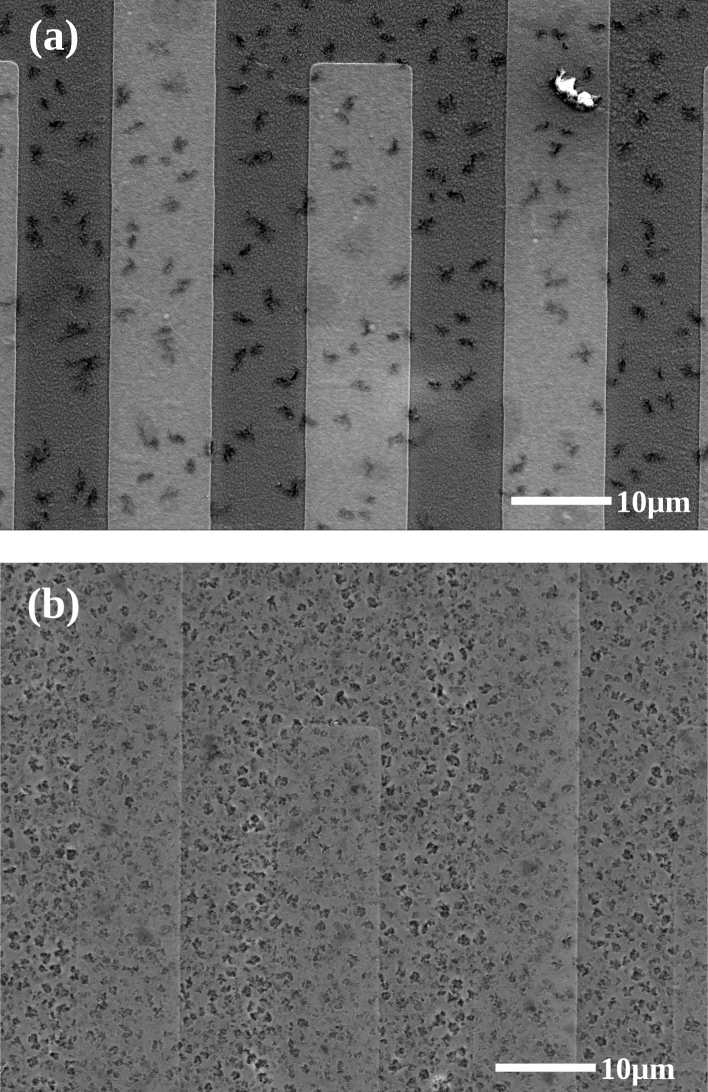}
    
    \caption{Scanning electron microscope (SEM) images of drop-cast NC films on the active areas of MSM photodetectors. Images show the distribution of (a) Cs$_y$WO$_{3-x}$ NCs and (b) WO$_{3-x}$ NCs. The solution concentrations were similar to those used in the main study. Both samples were loaded into the SEM at the same time and imaged under identical conditions (beam current, acceleration voltage, focus, and contrast settings).}\label{supp-fig:sem}
\end{figure}

\begin{figure}[H]
    \centering
    \includegraphics[width=\linewidth]{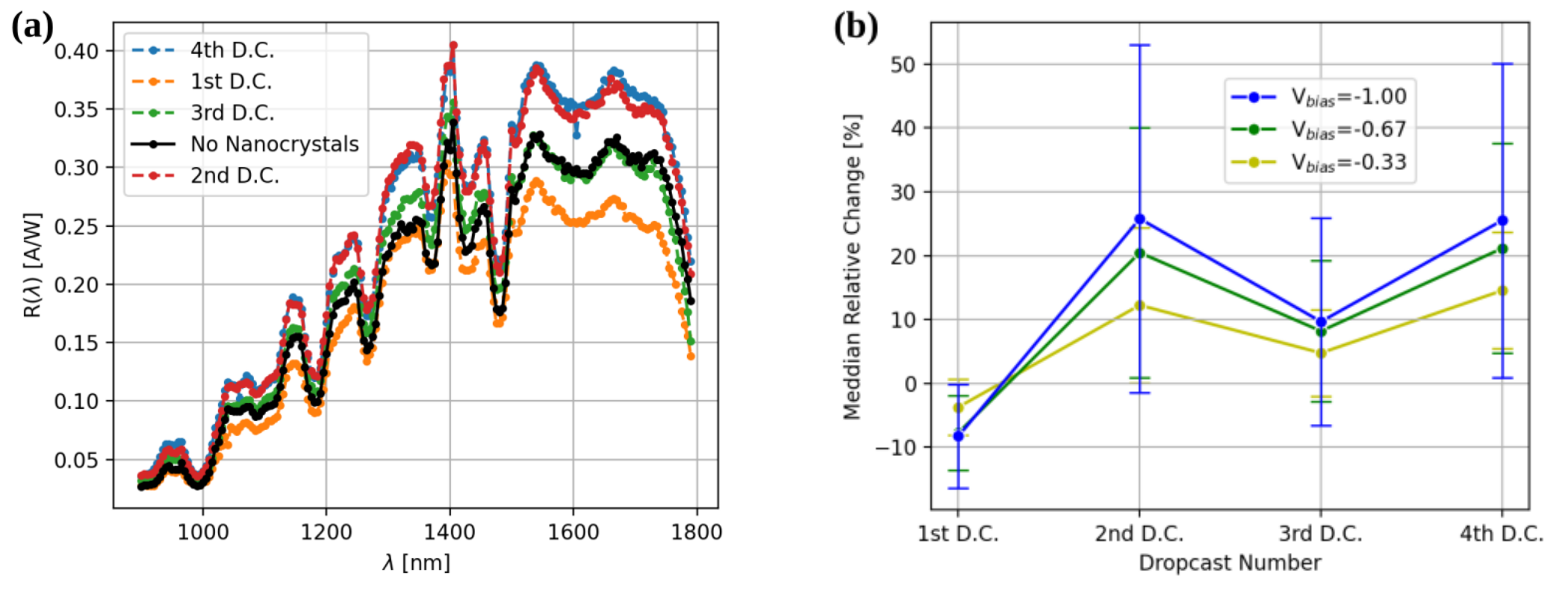}
    \caption{The effect of successive drop-casts of Cs$_y$WO$_{3-x}$ NCs using a concentration of 135g/L. (a) Responsivity as a function of wavelength for successive drop-casts. (b) Relative percent difference in responsivity for successive drop-casts, measured at several bias voltages. The median (data points) and standard deviation (bars) of the relative percent difference are calculated over the full wavelength range shown in (a).}\label{supp-fig:many_dropcasts}
\end{figure}

\begin{figure}[H]
    \centering
    \includegraphics[width=0.75\linewidth]{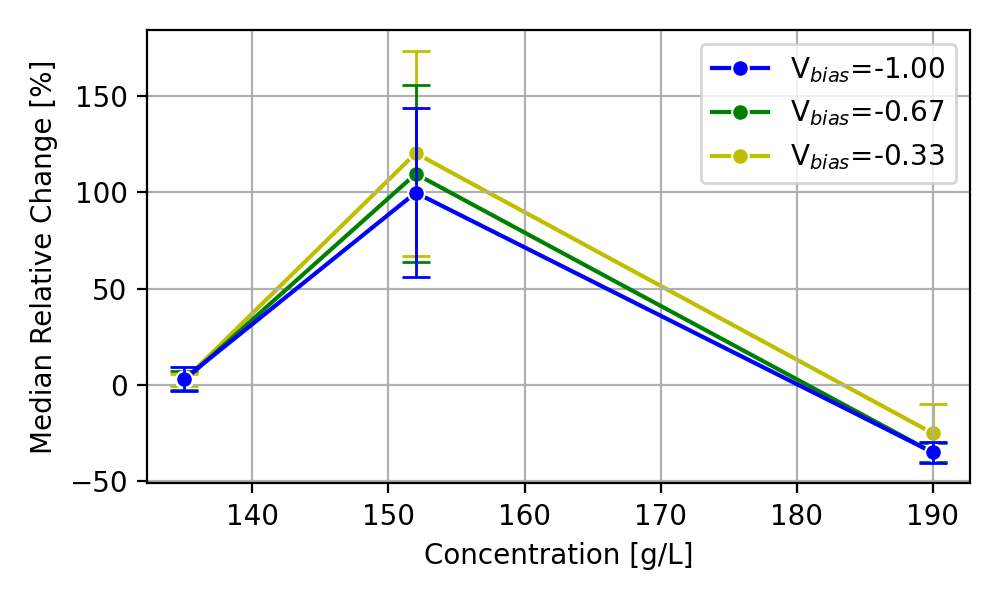}
    \caption{The effect of drop-cast concentration on the change in responsivity (RPD) for Cs$_y$WO$_{3-x}$ NCs. Samples were diluted in hexane. The RPD is measured over the full spectrum (900-1800 nm) and the median value plotted (bars indicate the standard deviation).} \label{supp-fig:dilution}
\end{figure}

A range of photodetector parameters were tested in various combinations creating 4 sets of photodetectors with the same geometries, L $\in \{400 \ \mu m, \ 500 \ \mu m\}$, S $\in \{3 \  \mu m, \ 10 \ \mu m\}$, and A $\in \{400 \ \mu m \times 440 \ \mu m, \ 500 \  \mu m \times 540 \  \mu m, \ 980 \  \mu m \times 540 \  \mu m\}$. Selected photodetectors were used for characterizing the quality and variability of drop-cast NCs as seen in figures: \ref{supp-fig:dropcast_worked}, \ref{supp-fig:dropcast_not_worked}, \ref{supp-fig:sem}, \ref{supp-fig:many_dropcasts}, and \ref{supp-fig:dilution}.

\end{document}